\documentclass[%
 reprint,
superscriptaddress,
 amsmath,amssymb,
 aps,
nofootinbib,
floatfix,
]{revtex4-2}

\usepackage{filecontents}
\usepackage{booktabs}

\usepackage{siunitx} %used for the math notation
\usepackage{xparse}
 \usepackage{url}

\usepackage{amsmath}
\usepackage[linesnumbered]{algorithm2e}% for algorithm notation

\usepackage{graphicx}% Include figure files
\usepackage{dcolumn}% Align table columns on decimal point
\usepackage{bm}% bold math
\usepackage{hyperref}% add hypertext capabilities
 \hypersetup{
     colorlinks=true,
     linkcolor=blue,
     filecolor=blue,
     citecolor =blue,      
     urlcolor=black,
     }

\begin{document}

\renewcommand*{\listalgorithmcfname}{List of Procedures}
\renewcommand*{\algorithmcfname}{Procedure}
\renewcommand*{\algorithmautorefname}{procedure}

\setlength\parskip{0pt}

\preprint{APS/123-QED}

%\title{Inverse variational autoencoder approach to the calculation of atomic-scale properties of chemically disordered materials}

\title{Targeting the partition function of chemically disordered materials with a generative approach based on inverse variational autoencoders}

\author{Maciej J. Karcz}
 %\email{maciej.karcz@cea.fr}
\affiliation{%
 CEA, DES, IRESNE, DEC, Cadarache, F-13108 Saint-Paul-Lez-Durance, France
}% 
\affiliation{
 Universit\'e Paris-Saclay, CEA, LIST, F-91120, Palaiseau France
}%
\author{Luca Messina}%
\email{luca.messina@cea.fr}
\affiliation{%
 CEA, DES, IRESNE, DEC, Cadarache, F-13108 Saint-Paul-Lez-Durance, France
}%

\author{Eiji Kawasaki}
\affiliation{
 Universit\'e Paris-Saclay, CEA, LIST, F-91120, Palaiseau France
}% 

\author{Emeric Bourasseau}%
 \affiliation{%
 CEA, DES, IRESNE, DEC, Cadarache, F-13108 Saint-Paul-Lez-Durance, France
}%

\date{\today}% It is always \today, today,

\begin{abstract}

Computing atomic-scale properties of chemically disordered materials requires an efficient exploration of their vast configuration space. Traditional approaches such as Monte Carlo or Special Quasirandom Structures either entail sampling an excessive amount of configurations or do not ensure that the configuration space has been properly covered. In this work, we propose a novel approach where generative machine learning is used to yield a representative set of configurations for accurate property evaluation and provide accurate estimations of atomic-scale properties with minimal computational cost. Our method employs a specific type of variational autoencoder with inverse roles for the encoder and decoder, enabling the application of an unsupervised active learning scheme that does not require any initial training database. The model iteratively generates configuration batches, whose properties are computed with conventional atomic-scale methods. These results are then fed back into the model to estimate the partition function, repeating the process until convergence. We illustrate our approach by computing point-defect formation energies and concentrations in $\mathrm{(U, Pu)O_2}$ mixed-oxide fuels. In addition, the ML model provides valuable insights into the physical factors influencing the target property. Our method is generally applicable to explore other properties, such as atomic-scale diffusion coefficients, in ideally or non-ideally disordered materials like high-entropy alloys.

% \begin{description}
% Secondary publications and information retrieval purposes.
% You may use the \texttt{description} environment to structure your abstract;
% use the optional argument of the \verb+\item+ command to give the category of each item. 
% \end{description}
\end{abstract}

\maketitle

%\tableofcontents

\section{\label{sec:Introcudtion}Introduction}

Chemical disorder presents a significant challenge in characterizing multi-component compounds. With various atomic species occupying the same crystal lattice, their distribution across lattice sites can vary widely, as seen in materials like high-entropy alloys (HEAs) \cite{lin2022defect, pickering2021high, george2019high} or mixed-actinide oxides used as nuclear fuels \cite{beauvy2009nuclear}. The vast number of potential configurations makes it computationally expensive to fully explore the configuration space needed to compute properties that depend on this distribution. Consequently, accurately characterizing these properties becomes a complex and demanding task.

One such property is the behavior of point defects, which provides essential insights into the atomic transport properties that influence the evolution of a material's microstructure from manufacturing through operational use. For instance, understanding the concentration of thermal defects is crucial for evaluating the material's fragility. However, due to the vast configuration space, access to this propertry is limited. A very common approach to the calculation of properties of chemically disordered solid solutions is the use of special quasirandom structures (SQS) \cite{zunger1990special, articleCyril}. Using SQS entails that the most disordered structures are expected to be the most probable ones, which is true for ideally disordered solutions, i.e., solutions with negligible mixing enthalpy. However, if the solution is not perfectly ideal, the most disordered structures are not necessarily the most probable ones. Even in perfectly disordered solutions, SQS generates only a limited number of structures, raising the question of whether these are adequate to comprehensively explore the configuration space for accurate property determination. Chemical disorder can also be approached using Markov Chain Monte Carlo (MCMC) techniques. For instance, in the work of Takoukam-Takoundjou \textit{et~al.} \cite{articleCyril}, MCMC is applied to compute the average thermodynamic properties of $\mathrm{(U, Pu)O_2}$ over a large number of sampled configurations. However, due to the large size of the configuration space, ensuring that the MCMC calculations explore it thoroughly demands significant computational resources \cite{cho2018sampling}. Another approach is to attempt a systematic exploration of the configuration space starting from a small system and progressively increasing its size, as shown in the study of Bathellier \textit{et~al.} \cite{didier2022disorder} on the defect formation energy in $\mathrm{(U, Pu)O_2}$.
%using an empirical interatomic potential \cite{cooper2014many}.
Nevertheless, such a method quickly becomes impractical, due to the exponentially increasing computational cost.

Previous works in the literature on the properties of chemically disordered solid solutions were restricted to the study of either very few configurations or relatively small systems. However, an alternative approach that could go beyond those limitations can be proposed with machine learning (ML). ML has spread out as an interdisciplinary field in various domains. It has been proven as a useful tool in constructing ML potentials \cite{dubois2024atomistic, hong2020training, andolina2021robust, dragoni2018achieving}, accelerating the modeling of battery materials \cite{guo2021accelerated, allam2018application}, augmenting the exploration of material properties \cite{kobayashi2022machine, ihalage2021analogical}, among many other applications. Exploration of the configuration space requires a method that is sufficiently efficient so it can be easily applied to changing environments and different properties. In the search for an approach that could allow for such exploration, we focus in this work on the bound Schottky defects (BSD) properties in $\mathrm{(U, Pu)O_2}$: namely the calculation of their formation energy and equilibrium concentration. These properties are crucial for characterizing radiation damage \cite{wiktor2014coupled, gautam2015static} and the fission-gas release process \cite{vathonne2017determination, thompson2011first}. The study of these properties is even more important due to the scarcity of data on BSD and other point defects in $\mathrm{(U, Pu)O_2}$ in the literature. 

\begin{figure}[t]
    \vspace{0cm} %\vspace{0 cm}
    %\setcapindent{0em}
    \centering
    \includegraphics[width=0.48\textwidth]{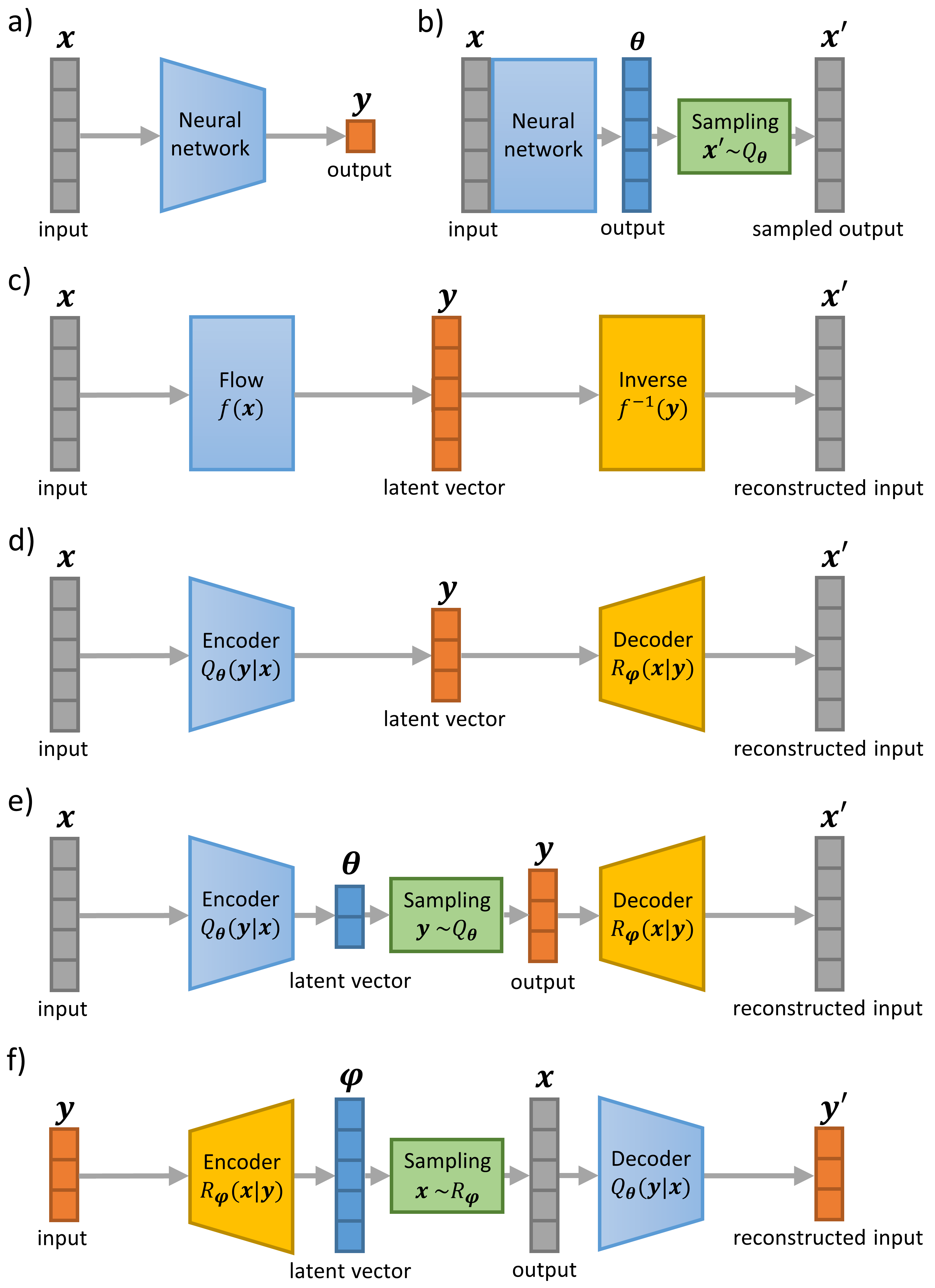}
    % First goest to the list of figures. Second is the description underneath
    \caption[Schematic representation of different machine learning models]{Schematic representation of various machine learning models: a) Classical feedforward neural network; b) Variational autoregressive network architecture from \cite{wu2019solving}. The model trains by sampling its output distribution; c) Normalizing flow architecture. The major difference from variational autoencoders comes from using invertible functions to map input data to latent representation, requiring the dimensions of both $\bm{x}$ and $\bm{y}$ to match; d) Autoencoder architecture; e) Variational autoencoder architecture, where $\bm{y}$ is sampled from the encoder distribution instead of being directly transformed from $\bm{x}$; f) Inverse variational autoencoder architecture, similar to classical variational autoencoders but with the roles of encoder and decoder reversed. Here, the dimensions of input and output do not need to match, allowing the sampling of high-dimensional $\bm{x}$ by sampling low-dimensional $\bm{y}$ from, for example, a Bernoulli distribution. Figure inspired by \cite{al2022qualitative, normalizingXLippe, anwar2021difference}. 
    }
    \label{fig:DifferentArchitecturesNfAeVaeIvae}
    \vspace{-0.2cm}
\end{figure}

Therefore, the goal of this work is to design a new approach to study the properties of chemically disordered compounds, such as uranium-plutonium mixed-oxide nuclear fuels, aiming for efficiency to save computational time. Since the more conventional approaches can prove this task to be very challenging \cite{didier2022disorder, articleCyril, njifon2018electronic}, we use ML methods to meet our objective. A specific type of machine learning that can be especially useful in our case is generative machine learning \cite{harshvardhan2020comprehensive}. These methods enable the generation of new data points based on the model's learned information about the environment, which can prove useful in our task of exploring the configuration space. The advantage of generative methods over more common applications of ML is the possible gain in computational time. In previous work \cite{karcz2023semi} we have introduced a generative semi-supervised approach based on Mixture Density Network (MDN) \cite{bishop1994mixture}. We demonstrated that evaluating and investigating local-atomic dependent properties in chemically disordered compounds can be efficiently achieved given appropriate data and information about the targeted system. However, such data is not always available or can be very expensive to obtain, which motivated us to explore alternative approaches.  What we need specifically is an approach capable of autonomously generating necessary data without initial inputs, enabling the approximation of the true distribution and efficient computation of targeted properties. This method should efficiently sample the configuration space, allowing us to compute these properties with minimal computational expense. In other words, our goal is to find a method to compute the system's partition function.

One interesting approach is using the Variational Inference (VI) method. VI aims to find a simpler distribution that is close to the true distribution, by turning the problem into an optimization task: minimizing the divergence between the true distribution and the approximate distribution. VI provides a training framework that can be used to train various machine learning models. One such method is normalizing flows \cite{rezende2015variational}. They allow to transform a simple distribution, such as a standard Gaussian distribution, into a more complex distribution through a sequence of invertible transformations. However, two problems arise. Firstly, normalizing flows uses the Jacobian change-of-variable transformations, whose computation and their determinants can become computationally expensive, especially for deep or complex flows. Secondly, because of used Jacobians, every step in the neural networks has to be invertible and they are not immediately applicable to discrete distributions. In our case, it is crucial as we focus on the discrete probability distribution of configurations where each configuration corresponds to a specific arrangement of atoms corresponding to their rigid-lattice positions. There are strategies to circumvent this issue and adapt the normalizing flows to the discrete scenarios either by the change of variables formula with the analogous normalizing flow setup \cite{tran2019discrete}, or by embedding categorical input data into $\mathcal{R}^k$ using variational inference in the first layer, and employing a continuous-data normalizing flow in the remaining layers \cite{lippe2020categorical}. Another approach is to train two neural networks using normalizing flows \cite{altosaar2020probabilistic}. The different architectures of the machine learning models discussed here and later in this section are visualized in Fig.~\ref{fig:DifferentArchitecturesNfAeVaeIvae}.

Variational autoregressive networks \cite{wu2019solving} are another example of machine learning models that can be trained within the VI framework to compute the probability distribution of a target system. They model the joint probability of all variables as the product of conditional probabilities parameterized by neural networks. Both of these approaches, whether using normalizing flows or autoregressive networks, are employed in machine learning to model complex probability distributions.

Another way to approach our objective could also be done through a different solution based on autoencoders \cite{gauerhof2020reverse, cantwell2022approximate}, which we refer to here as an Inverse Variational Autoencoder (IVAE). It is conceptually similar to Variational Autoencoder (VAE) \cite{ding2020guided, vahdat2020nvae, doersch2016tutorial}; however, VAEs in their common application still rely on data to train their encoder and decoder parts. To overcome that, in Inverse Variational Autoencoders (IVAEs), the roles of encoder and decoder are inversed. The encoder maps an input distribution that is simple and easy to sample (e.g. Gaussian or Bernoulli distribution) to a more complex one, while the decoder component reconstructs the original input distribution from this complex representation. Since input distribution is easy to sample, there is no need for an initial training database, allowing the model to essentially train on the data produced by itself. The process begins with sampling data from the input distribution, which is then transformed into a complex distribution. In our scenario, this complex distribution represents the distribution of atomic configurations in $\mathrm{(U, Pu)O_2}$. Subsequently, the decoder reconstructs the original distribution. 

While both normalizing flows and autoregressive models could be applied to our case, we decided to use IVAE because it offers more flexibility in its application compared to the other methods. Firstly, its input and output dimensions do not need to match, allowing us to use simple low-dimensional input distributions like Bernoulli to fit high-dimensional target distributions, potentially reducing computational costs. This is in contrast to normalizing flows, where input and output dimensions must match. Secondly, it can use more expressive forms of its output probability distribution compared to the variational autoregressive model from \cite{wu2019solving}, which requires its output distribution to be tractable. In contrast, IVAE does not have this constraint, possibly making it a better choice for approximating the Boltzmann distribution that we target in this work.

The properties of IVAE will be further elaborated in the following sections, where we demonstrate its performance in the real system of chemically disordered $\mathrm{(U, Pu)O_2}$. First, we will define the objective, which is to compute the point defect properties, specifically their formation energy and thermal-equilibrium concentration.  Next, we will explain how these properties can be determined by computing the partition function and discuss the mathematical formalism of IVAE models. Finally, we will present the results of applying IVAE to calculate the point defect properties.

\section{\label{sec:Methods}Mathematical framework}

\subsection{Concentration of thermal defects}

The concentration of vacancy-type defects can be defined \cite{karcz2023semi} as:

\begin{equation}\label{eq:CdT}
         C_{\mathrm{BSD}}(T) = \sum_{\bm{x}_\mathrm{c} \in \bm{X}} w(\bm{x}_\mathrm{c}) \exp\left({-\frac{E_{\mathrm{BSD}}^{\mathrm{f}}(\bm{x}_\mathrm{c})}{k_{\mathrm{B}}T}}\right).
\end{equation}

\noindent
$\bm{x}_\mathrm{c}$ corresponds to atomic configurations $\bm{x}_\mathrm{c} \in \bm{X}$; namely, a distribution of U/Pu atoms on the cationic sublattice - where $\bm{X} \in \mathbb{R}^{3N}$ is the space of atomic configurations. To be more precise, we consider this space $\bm{X}$ to consist of all possible $\bm{x}_\mathrm{c}$ where all degenerate symmetrically equivalent configurations are counted individually, ensuring that the multiplicity of each configuration is strictly equal to 1. As we focus on the BSD, $c$ denotes the type of cation to be removed from the supercell to form the Schottky defect.
$E_{\mathrm{BSD}}^{\mathrm{f}}(\bm{x}_\mathrm{c})$ is the formation energy of a vacancy type of Schottky defect and the sum in Eq.~(\ref{eq:CdT}) goes over all $\bm{x}_\mathrm{c}$ in $\bm{X}$. Here $w(\bm{x}_\mathrm{c})$ is the normalized weight that corresponds to %the Gibbs-Boltzmann distribution, i.e.
the probability of configuration $\bm{x}_\mathrm{c}$.
%and depends on the configuration energy $E_{\mathrm{conf}}(\bm{x}_\mathrm{c})$ as follows:

In the general case of a non-ideal solid solution, the BSD formation energy depends on the U/Pu chemical potentials that need to be determined. A recent work by Schuler et al. \cite{schuler2024towards, livacancy} has introduced a new method that allows for the determination of chemical potentials in the presence of short-range order based on the Widom substitution technique. In our work, we restrict the method to ideal solid solutions \cite{soisson2022atomistic, didier2022disorder}, leaving the generalization to non-ideal solutions in future work. 

In the case of an ideal solid solution, and assuming that we are removing U to form a Schottky defect, we can define $E_{\mathrm{BSD}}^{\mathrm{f}}(\bm{x}_\mathrm{c})$ as \cite{karcz2023semi}:

\begin{align}
    \label{eq:efbsdxcA}
    E_{\mathrm{BSD}}^{\mathrm{f}}(\bm{x}_\mathrm{U}) = E_{\mathrm{BSD}}(\bm{x}_\mathrm{U}) - E_{\mathrm{conf}}(\bm{x}_\mathrm{U}) + \frac{E(\mathrm{UO_2})}{N_\mathrm{C}}.
\end{align}

\noindent
$E_{\mathrm{BSD}}(\bm{x}_\mathrm{U})$ and $E_{\mathrm{conf}}(\bm{x}_\mathrm{U})$ represent the energies of a configuration $\bm{x}_\mathrm{U}$ with and without the removal of cation U and two oxygen atoms, respectively. $E(\mathrm{UO_2})$ is the energy reference of the pure $\mathrm{UO_2}$ supercell and $N_\mathrm{C}$ is the number of cations in the supercell. The Schottky defect could also be formed by removing a Pu cation instead. However, as shown in \cite{karcz2023semi}, the difference between the two approaches is negligible for ideal solid solutions such as the $\mathrm{(U, Pu)O_2}$ system as described by the Cooper-Rushton-Grimes (CRG) potential \cite{cooper2014many}.

%the difference introduced by this substitution is minimal according to the interatomic potential used (the Cooper-Rushton-Grimes or CRG potential \cite{cooper2014many}).

In an ideal solid solution,
%the probability of a given atomic composition 
the probability $p(n_{\mathrm{Pu}})$ of having a given number of Pu atoms in an atomic configuration (representing the local environment around the defect) can be described using
%is not influenced by local order and can be described using 
the binomial distribution formula:

\begin{equation}\label{eq:global_local_concentration}
        p(n_{\mathrm{Pu}})= \binom{M_\mathrm{C}} {n_{\mathrm{Pu}}} \hspace{2pt} y_{\mathrm{Pu}}^{n_{\mathrm{Pu}}} \hspace{2pt} (1 - y_{\mathrm{Pu}})^{n_{\mathrm{U}}}.
\end{equation}

\noindent
$M_\mathrm{C}$ is the amount of cations in the considered configuration, while $n_{\mathrm{Pu}}$ and $n_{\mathrm{U}}$ describe the amount of Pu and U respectively so that $M_\mathrm{C} = n_{\mathrm{U}} + n_{\mathrm{Pu}}$. Pu concentration is expressed with $y_{\mathrm{Pu}}$ and $y_{\mathrm{U}} = 1 - y_{\mathrm{Pu}}$. We can use Eq.(\ref{eq:global_local_concentration}) to define $w'(\bm{x}_\mathrm{c})$ and use it in place of $w(\bm{x}_\mathrm{c})$:

\begin{equation}\label{eq:p_prime_concentration}
        w'(\bm{x}_\mathrm{c}) = y_{\mathrm{Pu}}^{n_{\mathrm{Pu}}^{\bm{x}_\mathrm{c}}} \hspace{2pt} (1 - y_{\mathrm{Pu}})^{ n_{\mathrm{U}}^{\bm{x}_\mathrm{c}}}.
\end{equation}

\noindent
In Eq.~(\ref{eq:p_prime_concentration}) by $n_\mathrm{Pu}^{\bm{x}_\mathrm{c}}$, $n_\mathrm{Pu}^{\bm{x}_\mathrm{c}}$ we denote the amount of Pu, U within the atomic configuration $\bm{x}_\mathrm{c}$, so $M_\mathrm{C} = n_\mathrm{Pu}^{\bm{x}_\mathrm{c}} + n_\mathrm{U}^{\bm{x}_\mathrm{c}}$, and we can show that if we consider all $\bm{x}_\mathrm{c} \in \bm{X}$, then $\sum_{\bm{x}_\mathrm{c} \in \bm{X}} w'(n_\mathrm{Pu}^{\bm{x}_\mathrm{c}}) = 1$. Removing $\binom{M_\mathrm{C}} {n_{\mathrm{Pu}}}$ is necessary, because instead of iterating through all $n_\mathrm{Pu} \in [0, M_\mathrm{C}]$ , as in Eq.~(\ref{eq:global_local_concentration}), in Eq.~(\ref{eq:CdT}) we iterate through all $\bm{x}_\mathrm{c} \in \bm{X}$.

\subsection{Partition function of formation energy}

Our objective is to compute defect concentration from Eq.~(\ref{eq:CdT}). Let us define it in a way that will be more easily approachable from the IVAE perspective:

\begin{equation}\label{eq:CdT_concentration_compressed}
\begin{split}
    C_{\mathrm{BSD}}(T) &= \sum_{\bm{x}_\mathrm{c} \in \bm{X}}  w'(\bm{x}_\mathrm{c})  \exp\left({-\frac{E_{\mathrm{BSD}}^{\mathrm{f}}(\bm{x}_\mathrm{c})}{k_{\mathrm{B}}T}}\right)  \\
    &= \sum_{\bm{x}_\mathrm{c} \in \bm{X}} \exp\left({-\frac{E_{\mathrm{BSD}}^{\mathrm{f}}(\bm{x}_\mathrm{c})}{k_{\mathrm{B}}T}} + \ln{\left(w'(\bm{x}_\mathrm{c})\right)}  \right). \\
\end{split}
\end{equation}

\noindent
To simplify the notation we will denote the exponential term from Eq.~(\ref{eq:CdT_concentration_compressed})  by $f_T(\bm{x}_\mathrm{c})$, so for a given temperature $T$:

\begin{equation}\label{eq:f'x}
         f_T(\bm{x}_\mathrm{c}) = \exp\left({-\frac{E_{\mathrm{BSD}}^{\mathrm{f}}(\bm{x}_\mathrm{c})}{k_{\mathrm{B}}T}} + \ln{\left(w'(\bm{x}_\mathrm{c})\right)}  \right),
\end{equation}

\noindent
so:

\begin{equation}\label{eq:CdT_concentration_compressed_Zf}
                 C_{\mathrm{BSD}}(T) = \sum_{\bm{x}_\mathrm{c} \in \bm{X}} f_T(\bm{x}_\mathrm{c}) = Z_T.
\end{equation}

From now on we will call $Z_T$ as the "partition function of formation energy". Therefore, by approximating $Z_T$, with e.g. IVAE, we can directly approximate $ C_{\mathrm{BSD}}(T)$. It is important to note that we called $Z_T$ a partition function to facilitate the discussion in the further parts of this work and to show that the IVAE approach can be generally applied to the evaluation of any partition function of that form, but it does not correspond to the partition function describing the thermodynamical equilibrium of the system. Nonetheless, in principle, we can apply the discussed IVAE approach to target any partition function.

\subsection{Mathematical Formalism of IVAE}

Our goal, as stated in the previous section is to compute $Z_T$ - the partition function of formation energy. $Z_T$ is used to compute the probabilities of the states of the system and we can use it to define the probability of a given configuration $\bm{x}_\mathrm{c}$ in the following way:

\begin{equation}\label{eq:prob_of_xc_Z}
    P_{T}(\bm{x}_\mathrm{c}) = \frac{{f_T(\bm{x}_\mathrm{c})}}{{Z_T}}.
\end{equation}

\noindent
If we can determine $Z_T$, we can define $P_{T}(\bm{x}_\mathrm{c})$. Conversely, if we know $P_{T}(\bm{x}_\mathrm{c})$, we can compute $Z_T$. In most practical scenarios, $P_{T}(\bm{x}_\mathrm{c})$ is an unknown and potentially complex distribution of the target system, so trying to compute it directly is challenging. 

One approach is to find another distribution $P(\bm{y})$ and use it to obtain $P_{T}(\bm{x}_\mathrm{c})$. The idea is that if we choose $P(\bm{y})$ to be simple and easy to sample from and manage to connect or transform it to $P_{T}(\bm{x}_\mathrm{c})$, we could sample $\bm{y}$ from $P(\bm{y})$ to obtain $\bm{x}_\mathrm{c}$ distributed according to $P_{T}(\bm{x}_\mathrm{c})$, which will bring us closer to computing the partition function $Z_T$. To achieve this, we can define an arbitrary probability distribution $R(\bm{x}_\mathrm{c} | \bm{y})$, which we can use to sample $\bm{x}_\mathrm{c}$ given $\bm{y}$. However, $\bm{x}_\mathrm{c}$ obtained in this way is not yet distributed according to $P_{T}(\bm{x}_\mathrm{c})$, and highly depends on the form of $R(\bm{x}_\mathrm{c} | \bm{y})$.

To connect $R(\bm{x}_\mathrm{c} | \bm{y})$ to $P_{T}(\bm{x}_\mathrm{c})$, we can use Bayes' theorem for conditional distributions. To do so, we need to first define another conditional distribution $Q(\bm{y} | \bm{x}_\mathrm{c})$, that allows us to do the reverse - sample $\bm{y}$ given $\bm{x}_\mathrm{c}$. We assume $P(\bm{y})$, $R(\bm{x}_\mathrm{c} | \bm{y})$ and $Q(\bm{y} | \bm{x}_\mathrm{c})$ to be normalized, so $\sum_{\bm{y}}P(\bm{y}) = 1$, $\sum_{\bm{y}}Q(\bm{y}|\bm{x}_\mathrm{c}) = 1$ and
$\sum_{\bm{x}_\mathrm{c}}R(\bm{x}_\mathrm{c}|\bm{y}) = 1$. We can now write:

\begin{equation}\label{eq:px_eq_pyrq}
    P_{T}(\bm{x}_\mathrm{c})  = \frac{P(\bm{y}) R(\bm{x}_\mathrm{c} | \bm{y})}{Q(\bm{y} | \bm{x}_\mathrm{c})}. 
\end{equation}

Eq.~(\ref{eq:px_eq_pyrq}) presents an ideal case scenario, where R and Q distributions allow for an exact mapping between $ P_{T}(\bm{x}_\mathrm{c})$ and $P(\bm{y})$. Since, as discussed before, finding an exact solution can be difficult in practice, we aim to find such R and Q distributions, to approximate $P_{T}(\bm{x}_\mathrm{c})$ in Eq.~(\ref{eq:px_eq_pyrq}). Let us rewrite Eq.~(\ref{eq:px_eq_pyrq}) in a different form:

\begin{equation}\label{eq:pxq_eq_pyr}
            P_{T}(\bm{x}_\mathrm{c}) Q(\bm{y} | \bm{x}_\mathrm{c}) = P(\bm{y}) R(\bm{x}_\mathrm{c} | \bm{y}).
\end{equation}

Our objective is to turn Eq.~(\ref{eq:pxq_eq_pyr}) into an optimization problem, where we could parameterize both R and Q and turn the goal of approximating $P_{T}(\bm{x}_\mathrm{c})$ into finding the best set of parameters for R and Q distributions. Therefore, we need a way to quantify whether a given set of parameters performs better or worse. We can achieve this by measuring the difference between both sides of Eq.~(\ref{eq:pxq_eq_pyr}) using a standard approach based on the Kullback-Leibler (KL) divergence \cite{joyce2011kullback, ihalage2021analogical, brunton2022data}. Since we want to compute variational inference and thus do not rely on any initial training database, we apply the reverse KL divergence to Eq.~(\ref{eq:pxq_eq_pyr}) so it becomes:

\begin{equation}\label{eq:KL_pxq_eq_pyr}
    \begin{split}
    D_{KL}\left[ P(\bm{y}) R(\bm{x}_\mathrm{c} | \bm{y}) \; \parallel \; P_{T}(\bm{x}_\mathrm{c}) Q(\bm{y} | \bm{x}_\mathrm{c}) \right] 
    \\
        \begin{split}
                &=
        \mathbb{E}_{ \bm{y} \sim P, \bm{x}_\mathrm{c} \sim R} \ln \left[ \frac{P(\bm{y}) R(\bm{x}_\mathrm{c} | \bm{y})}{P_{T}(\bm{x}_\mathrm{c}) Q(\bm{y} | \bm{x}_\mathrm{c})}  \right]
        \\
        &=
        - \mathbb{E}_{ \bm{y} \sim P, \bm{x}_\mathrm{c} \sim R} \ln \left[ \frac{P_{T}(\bm{x}_\mathrm{c}) Q(\bm{y} | \bm{x}_\mathrm{c})}{P(\bm{y}) R(\bm{x}_\mathrm{c} | \bm{y})}  \right].
        \end{split}
    \end{split}
\end{equation}

Eq.~(\ref{eq:KL_pxq_eq_pyr}) provides a framework that allows us to measure how closely we approximate $P_{T}(\bm{x}_\mathrm{c})$, depending on the choices of R and Q distributions, with the objective of minimizing its right-hand side. We can now come back to our original task stated at the beginning of this section - computation of the partition function $Z_T$ - and show how we can achieve it using Eq.~(\ref{eq:KL_pxq_eq_pyr}). Because KL divergence is always greater or equal to 0, we can write:

\begin{equation}\label{eq:KL_always_positive}
    0 \leqslant - \mathbb{E}_{ \bm{y} \sim P, \bm{x}_\mathrm{c} \sim R} \ln \left[ \frac{P_{T}(\bm{x}_\mathrm{c}) Q(\bm{y} | \bm{x}_\mathrm{c})}{P(\bm{y}) R(\bm{x}_\mathrm{c} | \bm{y})}  \right].
\end{equation}

\noindent
We can now use Eq.~(\ref{eq:prob_of_xc_Z}) for $P_{T}(\bm{x}_\mathrm{c})$ in Eq.~(\ref{eq:KL_always_positive}) and change it in a following way:

\begin{equation}\label{eq:KL_always_positive}
    \begin{split}
    0 &\leqslant - \mathbb{E}_{ \bm{y} \sim P, \bm{x}_\mathrm{c} \sim R} \ln \left[ \frac{\frac{{f_T(\bm{x}_\mathrm{c})}}{{Z_T}} Q(\bm{y} | \bm{x}_\mathrm{c})}{P(\bm{y}) R(\bm{x}_\mathrm{c} | \bm{y})}  \right] \\
    0 &\geqslant \mathbb{E}_{ \bm{y} \sim P, \bm{x}_\mathrm{c} \sim R} \left[\ln \left[ \frac{  f_T(\bm{x}_\mathrm{c}) Q(\bm{y} | \bm{x}_\mathrm{c})}{P(\bm{y}) R(\bm{x}_\mathrm{c} | \bm{y})}  \right] - \ln{Z_T} \right], \\    
    \end{split}
\end{equation}

\noindent
so it becomes:

\begin{equation}\label{eq:lnZ_fq_Jensen}
    \ln{Z_T} \geqslant \mathbb{E}_{ \bm{y} \sim P, \bm{x}_\mathrm{c} \sim R} \ln \left[ \frac{  f_T(\bm{x}_\mathrm{c}) Q(\bm{y} | \bm{x}_\mathrm{c})}{P(\bm{y}) R(\bm{x}_\mathrm{c} | \bm{y})} \right].
\end{equation}

Because of the change of signs in Eq.~(\ref{eq:KL_always_positive}), to estimate $\ln{Z_T}$, we need to maximize the right-hand side of Eq.~(\ref{eq:lnZ_fq_Jensen}). This is equvalent to computing the best evidence lower bound (ELBO), with the unconstrained optimum expressed as in Eq.~(\ref{eq:px_eq_pyrq}), for which $\ln{Z_T}$ is computed exactly. As discussed before in Eq.~(\ref{eq:px_eq_pyrq}), it also allows us to sample $\bm{x}_\mathrm{c}$ from $P_{T}(\bm{x}_\mathrm{c})$. This involves initially sampling $\bm{y}$ from $P(\bm{y})$ and then sampling $\bm{x}_\mathrm{c}$ from $R(\bm{x}_\mathrm{c}|\bm{y})$, resulting in $\bm{x}_\mathrm{c}$ being distributed according to $P_{T}(\bm{x}_\mathrm{c})$. However, obtaining the optimum that maximizes Eq.~(\ref{eq:lnZ_fq_Jensen}) is challenging in practical scenarios. Thus, the objective is to identify useful forms of the $R$, $P(\bm{y})$ and $Q$ distributions, that could help us approximate $\ln{Z_T}$. 

To define $R$ we can use the product distribution proposed in \cite{cantwell2022approximate}, with slight notation adjustment as follows:

\begin{equation}\label{eq:Rxy_phi}
     R_{\bm{\phi}}(\bm{x}_\mathrm{c}|\bm{y}) = \prod_i \frac{e^{ { I(x^{(i)}_{\mathrm{c}}) \phi^{(i)}(\bm{y})}}}{e^{\phi^{(i)}(\bm{y})}+e^{-\phi^{(i)}(\bm{y})}}.
\end{equation}

\noindent
However, since in our case of $\mathrm{(U, Pu)O_2}$ each element $x^{(i)}_\mathrm{c}$ of the configuration $\bm{x}_\mathrm{c}$ corresponds to either Pu or U, we introduce an indicator function $I(x^{(i)}_{\mathrm{c}})$ that would assign values 1 or -1 to Pu or U elements of $\bm{x}_\mathrm{c}$ respectively. 

\begin{equation}\label{eq:indicator}
    I(x^{(i)}_{\mathrm{c}}) = \begin{cases}
    \phantom{-}1 & \text{if } x^{(i)}_{\mathrm{c}} \text{ represents Pu,}  \\
    -1 & \text{if } x^{(i)}_{\mathrm{c}} \text{ represents U.}
    \end{cases}
\end{equation}

The change of notation from $R$ to $R_{\bm{\phi}}$ serves two main purposes. The first is to easily distinguish $\phi$ as a parameter of $R$ distribution. The second is to denote that $R_{\bm{\phi}}$ describes a multidimensional distribution. The vector of parameters $\bm{\phi}$ consists of several elements $\phi^{(i)}(\bm{y})$ that now depend on the random variable $\bm{y}$. We represent $\bm{\phi}$ using a neural network, forming the encoder part of the IVAE architecture.

There is an intuitive way on how to interpret Eq.~(\ref{eq:Rxy_phi}). Each parameter $\phi^{(i)}(\bm{y})$ can be associated with the occupancy (U/Pu) of each atom contained in the generated configuration and the number of predicted elements  $\phi^{(i)}(\bm{y})$ is equal to the number of atoms in the atomic configuration. If during the network training, it turns out that some atomic positions are more likely to be Pu, then the corresponding $\phi^{(i)}(\bm{y})$ will increase to higher positive values. Otherwise, if U is more likely, then $\phi^{(i)}(\bm{y})$ will decrease to lower negative values. To obtain strict occupancy probabilities, a sigmoid function can be applied to $\phi^{(i)}(\bm{y})$. The same logic applies to the later defined Eq.~(\ref{eq:Qyx_theta}).

We are now able to move forward to the remaining distributions. A convenient choice for  $P(\bm{y})$ is to define $\bm{y}$ as a series of independent coin flips, with each $y^{(i)}$ having a probability of 1/2 of being either 1 or -1. For $Q$ distribution we can follow the same logic as in the case of $R_{\bm{\phi}}$ distribution in Eq.~(\ref{eq:Rxy_phi}), so after applying a similar change of notation it becomes:

\begin{equation}\label{eq:Qyx_theta}
     Q_{\bm{\theta}}(\bm{y}|\bm{x}_\mathrm{c}) = \prod_i \frac{e^{{y^{(i)} \theta^{(i)}(\bm{x}_\mathrm{c})}}}{e^{\theta^{(i)}(\bm{x}_\mathrm{c})}+e^{-\theta^{(i)}(\bm{x}_\mathrm{c})}}.
\end{equation}

\noindent
Similarly as in Eq.~(\ref{eq:Rxy_phi}), we will represent $\bm{\theta}$ with the neural network, that corresponds to the decoder part of IVAE. With the updated notation, we can rewrite Eq.~(\ref{eq:lnZ_fq_Jensen}):

\begin{equation}\label{eq:lnZ_fq_Jensen_loss}
    \ln{Z_T} \geqslant \mathbb{E}_{ \bm{x}_\mathrm{c} \sim R_{\bm{\phi}},\bm{y} \sim P}\left[\ln{\frac{f_T(\bm{x}_\mathrm{c}) Q_{\bm{\theta}}(\bm{y}|\bm{x}_\mathrm{c})}{ 
    P(\bm{y}) R_{\bm{\phi}}(\bm{x}_\mathrm{c} | \bm{y})   }  }\right] = -\Tilde{F}_{\bm{\phi}, \bm{\theta}},
\end{equation}

\noindent
where $-\Tilde{F}_{\bm{\phi}, \bm{\theta}}$ is the lower bound of $\ln{Z_T}$ that could be referred to as loss function of IVAE, which we will maximize to estimate $\ln{Z_T}$. 

\subsection{Unsupervised training loop and functionality of IVAE}

\begin{figure}[h!tbp]
    \vspace{0.5cm}
    \centering
    \includegraphics[width=0.48\textwidth]{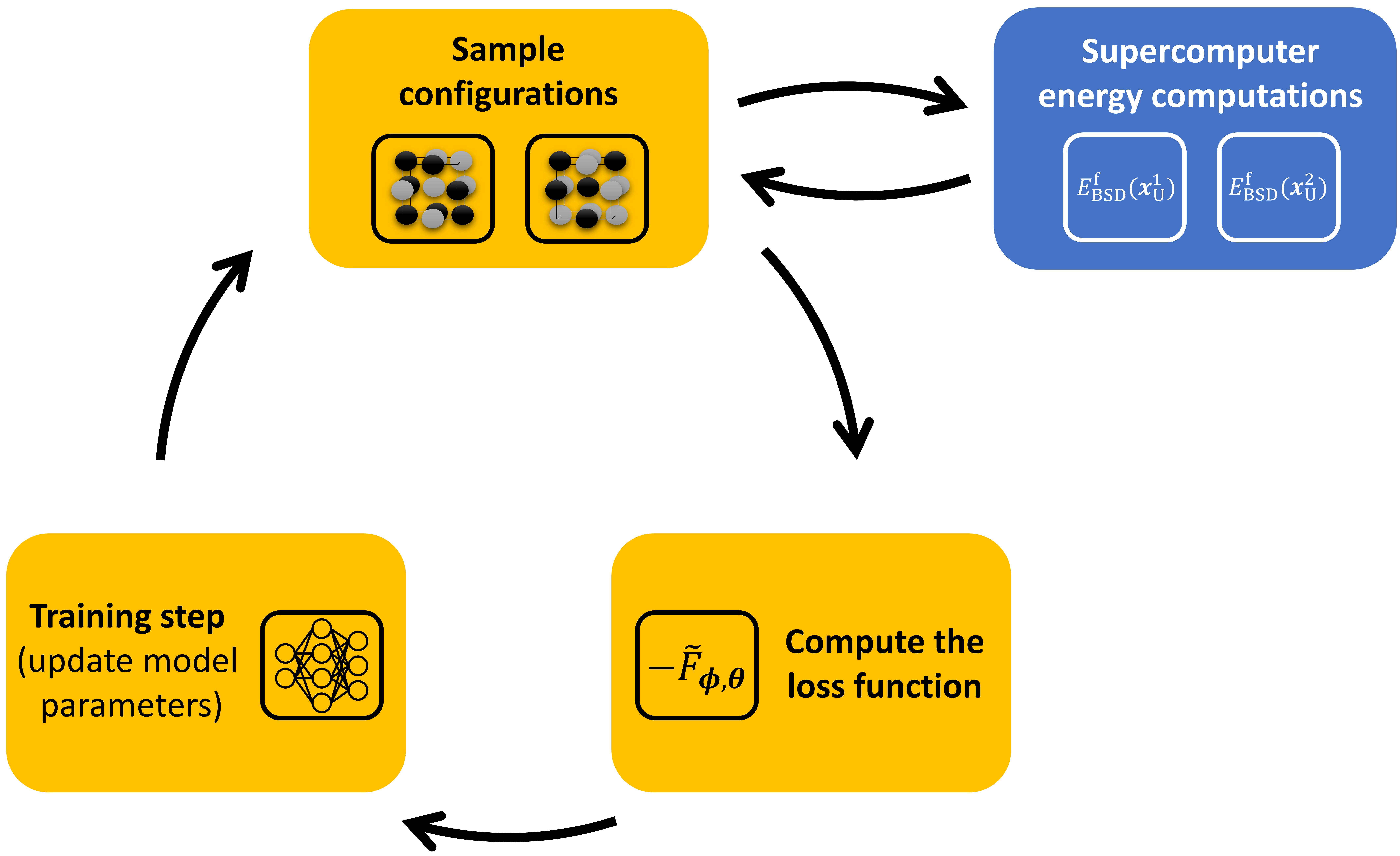}
    % First goest to the list of figures. Second is the description underneath
    \caption[Schematic visualization of IVAE model unsupervised training loop]{ Schematic visualization of the IVAE model's unsupervised training loop. Initially, the model parameters are randomly initialized. At each iteration, we sample $\bm{y}$ from $P(\bm{y})$ and then atomic configurations $\bm{x}_\mathrm{c}$ from $R_{\bm{\phi}}(\bm{x}_\mathrm{c}|\bm{y})$, compute the loss function, and update the model weights. This process is repeated until convergence.}
    \label{fig:TrainingLoopIVAE}
\end{figure}

\begin{figure*}[h!tbp]
    \vspace{0.5cm}
    \centering
    \includegraphics[width=0.9\textwidth]{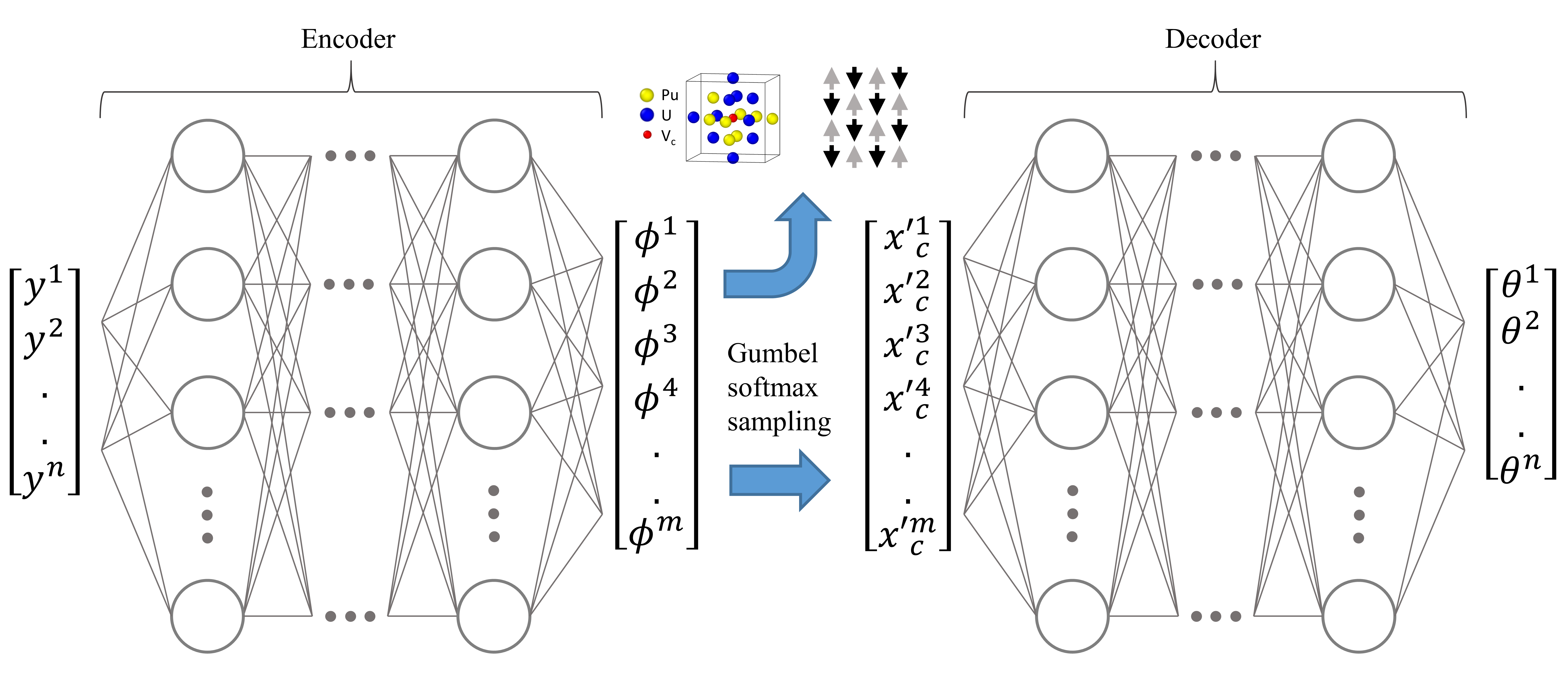}
    % First goest to the list of figures. Second is the description underneath
    \caption[General architecture of the used IVAE models]{General architecture of the used IVAE models. Samples from the input $P(\bm{y})$ distribution are represented with $y$. $\phi$ corresponds to the predicted parameters of $R_{\bm{\phi}}$ distribution from Eq.~(\ref{eq:Rxy_phi}). $x'_c$ are the samples from the Gumbel softmax distribution from the original $R_{\bm{\phi}}$ and $\theta$ are the predicted parameters of $Q_{\bm{\theta}}$ distribution from Eq.~(\ref{eq:Qyx_theta}). 
    After the model's training is finished, $R_{\bm{\phi}}$ can be used to, e.g. generate atomic configurations. In principle, the model can target different partition functions, thus allowing for sampling other distributions from different systems, e.g. Ising spin systems.}
    \label{fig:arch_IVAE}
\end{figure*}

Fig.~\ref{fig:TrainingLoopIVAE} shows a schematic visualization of the IVAE model's unsupervised training loop and Fig.~\ref{fig:arch_IVAE} presents the general architecture of the IVAE model presented so far. Let's summarize the entire process discussed so far. Our objective is to maximize the right-hand side of (\ref{eq:lnZ_fq_Jensen_loss}), which serves as the lower bound of our targeted partition function and acts as the loss function for IVAE. This formulation ensures that any prediction made by the IVAE model will always be lower or equal to the true partition function of the targeted system and facilitates the training process, by transforming it into an optimization task.

IVAE undergoes training in an unsupervised loop. In each iteration, it samples $\bm{y}$ from $P(\bm{y})$ and then samples atomic configurations $\bm{x}_\mathrm{c}$ from $R_{\bm{\phi}}(\bm{x}_\mathrm{c} | \bm{y})$. Since $\bm{x}_\mathrm{c}$ is discrete, the Gumbel softmax trick \cite{jang2016categorical, sun2021novel} is applied. This technique uses Gumbel-distributed noise to approximate the sampling of $\bm{x}_\mathrm{c}$ from $R_{\bm{\phi}}(\bm{x}_\mathrm{c} | \bm{y})$ with a softmax function. This enables gradient-based optimization through the categorical sampling process, allowing backpropagation to train the model effectively.

%Formation energies are computed for each group of generated configurations externally. In each training iteration, atomic configurations are exported to the Joliot-Curie supercomputer for energy minimization using classical molecular statics (or 0-K energy minimization) with the LAMMPS code \cite{LAMMPS} and the Cooper-Rushton-Grimes (CRG) empirical potential \cite{cooper2014many}. Once the formation energies are calculated, they are imported back into the model to compute its loss function, update the model's parameters, and advance to the next iteration of the training loop.

We train two networks simultaneously. The first plays the role of an encoder that predicts $\bm{\phi}$ parameters of $R_{\bm{\phi}}(\bm{x}_\mathrm{c} | \bm{y})$, transforming auxiliary variables $\bm{y}$ into configurations $\bm{x}_\mathrm{c}$. The second is predicting the parameters $\bm{\theta}$ of $Q_{\bm{\theta}}(\bm{y}|\bm{x}_\mathrm{c})$, in order to decode sampled $\bm{x}_\mathrm{c}$ back to $\bm{y}$. This is similar to the classical implementation of VAE, but with inverse roles of encoder and decoder, as discussed before. We encode samples $\bm{y}$ from a traceable input distribution $P(\bm{y})$ into the space of atomic configurations approximated with $R_{\bm{\phi}}(\bm{x}_\mathrm{c} | \bm{y})$ and then we decode it back. The big advantage over the VAE models is that IVAE does not need to be trained on some pre-prepared data of $\bm{x}_\mathrm{c}$ from the target distribution $P_{T}(\bm{x}_\mathrm{c})$, which it generates autonomously during training.

%While formation energy part of $f_T(\bm{x}_\mathrm{c})$ is fairly typical, there is an interesting interpretation of $\ln{\left(w'(\bm{x}_\mathrm{c})\right)}$ part of Eq.~(\ref{eq:f'x}). The configuration probability $w'(\bm{x}_\mathrm{c})$, as defined in Eq.~(\ref{eq:p_prime_concentration}), acts somewhat like a 'magnetic field' that favors configurations with generated Pu content aligning with the predefined $y_{\mathrm{Pu}}$ Pu concentration.

During each iteration of the training loop, where we sample $M$ variables $\bm{y}$ and configurations $\bm{x}_\mathrm{c}$, we estimate $-\Tilde{F}_{\bm{\phi, \theta}}$ using the following formula:

\begin{equation}\label{eq:ELBO_MC}
    -\Tilde{F}_{\bm{\phi, \theta}} \simeq \frac{1}{M}\sum_{i=1}^{M} \ln{\frac{f_T(\bm{x}^{(i)}_{\mathrm{c}}) Q_{\bm{\theta}}(\bm{y}^{(i)}|\bm{x}^{(i)}_{\mathrm{c}})}{ 
    P(\bm{y}^{(i)}) R_{\bm{\phi}}(\bm{x}^{(i)}_{\mathrm{c}} | \bm{y}^{(i)})   }  }.
\end{equation}

\noindent
$M$ is one of the hyperparameters of the model. The sampling process can be customized in various ways, but in our implementation, we sampled one $\bm{x}_\mathrm{c}$ for each sampled $\bm{y}$, as it was giving satisfactory results.

\section{Results}

In this section, we apply the IVAE model to the target MOX system. First, we test and optimize the model by applying it to smaller atomic environments. These smaller systems allow us to easily compute the true partition function, which we can then compare with the model predictions to assess its accuracy. To additionally verify IVAE performance, we also apply it to the Ising system in Appendix \ref{sec:AppendixIsing}, reproducing the results obtained in \cite{cantwell2022approximate}. After this initial testing and optimization, we proceed to apply the IVAE model to larger atomic environments.

\subsection{Description of the atomic environment in MOX with IVAE}\label{sec:IVAE_sphere_external_environment}

% Our goal is to compute the defect concentration in $\mathrm{(U, Pu)O_2}$, taking the bound Schottky defect as a test case, which can be achieved, as previously shown in Eq.~(\ref{eq:CdT_concentration_compressed_Zf}), by calculating the partition function of the formation energies. Since computing the partition function exactly is nearly impossible, we use IVAE to approximate it. The loss function of the IVAE model, described in Eq.~(\ref{eq:lnZ_fq_Jensen_loss}), directly corresponds to estimating the true partition function and is designed to be its lower bound. Therefore, we aim to maximize the value of the IVAE loss function. IVAE trains iteratively by generating a batch of configurations, computing their energy, and making estimations of the loss function as shown in Eq.~(\ref{eq:ELBO_MC}). Once the training is complete, the newly generated samples of configurations should approximately follow the distribution $P(\bm{x}_\mathrm{c})$.

In our application, generating configurations involves two steps. Firstly, as we focus on the bound Schottky defect properties, we generate configurations containing atoms in the closest vicinity of the defect, called nearest neighbor (nn) spheres of configurations or spheres of influence. This limits the configuration space to possible atomic configurations on the $x$nn sphere, where $x$ varies depending on the sphere radius. As we observed previously with the MDN approach and will see in subsequent sections, this space reduction, while being an approximation, is sufficient to obtain satisfactory results and potentially reduce computational time. Secondly, we construct the external environment around the generated spheres and apply periodic boundary conditions. If the system is too small, there's a risk of atoms interacting with their periodic images. Therefore we perform energy minimization on atomic configurations built as 2592-atom supercells (864 cations and 1728 anions) of $\mathrm{(U, Pu)O_2}$ with a 50\% Pu concentration. In the case of MOX, this corresponds to a $6 \times 6 \times 6$ replication of a primitive cell of a fluorite structure. 

The external environment for each generated sphere is prepared in two ways. The first method uses a reference supercell, where we replace its central atoms with those generated by the IVAE model, following the procedure described in \cite{karcz2023semi}. This keeps the external environment around the generated spheres unchanged across experiments but may introduce slight variations in the total Pu concentration since the IVAE model is not constrained to place an exact number of Pu atoms within generated configurations. The second approach involves randomly filling the environment around nn spheres to achieve the desired global Pu concentration. For each generated sphere, we count the number of placed Pu atoms and then randomly fill the rest of the supercell to attain the targeted total number of Pu atoms in the supercell.

In the study of the range of influence presented later on, we combine both approaches. We first generate an nn sphere with IVAE, then add a random environment around it, and finally place this prepared configuration in a reference supercell by replacing its central atoms with those generated by the IVAE model, in a similar way to that that was described in \cite{karcz2023semi}. For each of the following experiments, we specify how the external environment was prepared: with the constant reference supercell, random external environment, or as a mixture of both.

From a computational perspective, the BSD formation energies in each configuration are obtained via classical molecular statics (or 0-K energy minimization) using the LAMMPS code \cite{LAMMPS} and the Cooper-Rushton-Grimes (CRG) interatomic potential \cite{cooper2014many}. As a reminder, the computation of the formation energy for each configuration involves energy minimization of two supercells: one with and one without a defect. All supercells with a defect were prepared by removing a U atom to form a BSD. The formula used for the formation energy calculation is described in Eq.~(\ref{eq:efbsdxcA}). 

\subsection{Optimizing IVAE training for varying nominal Pu concentrations}~\label{sec:optimizing_IVAE}

The IVAE model's effectiveness for the generation of atomic configurations relies on the predicted probability distribution $R_{\bm{\phi}}$. This distribution allows the model to determine the likelihood of having U or Pu at specific positions on the crystal lattice. The parameters $\bm{\phi}$, initially set randomly, are adjusted during training to optimize the loss function. Proper initialization of $\bm{\phi}$ can significantly reduce training times by starting closer to the optimal values.

\begin{figure}[h!tbp]
    \vspace{0.5cm}
    \centering
    \includegraphics[width=0.48\textwidth]{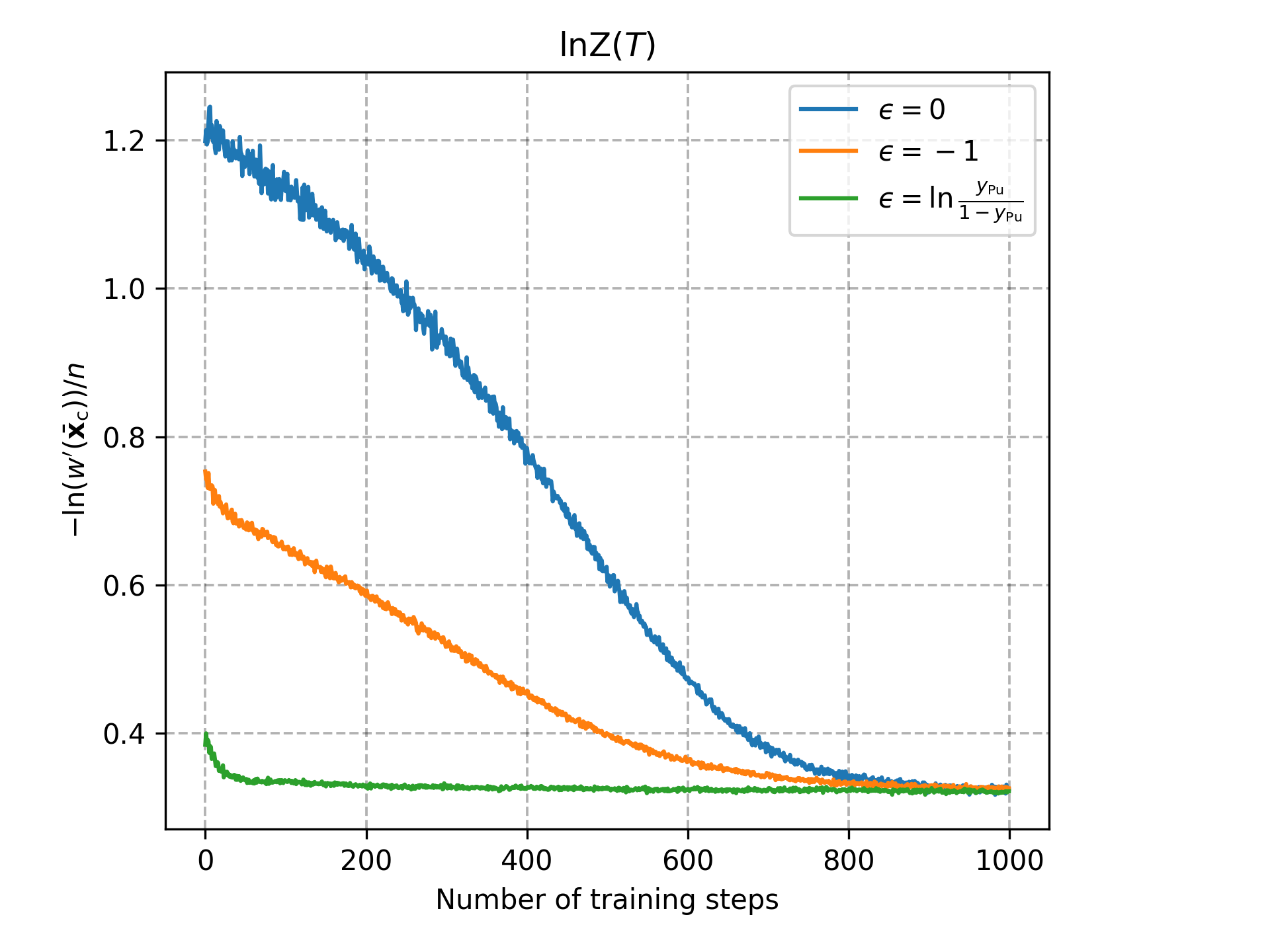} \hspace{30pt}
    % First goest to the list of figures. Second is the description underneath
    \caption[The training process of the configuration probability part of the IVAE lower bound]{The training process of the configuration probability part of the IVAE lower bound. The experiment was done for T = 500 K, $|\bm{y}| = 8$, for a global $y_{\mathrm{Pu}}$ concentration equal 10\%. $\bar{\bm{x}}_\mathrm{c}$ indicates that the configuration probability part of the IVAE lower bound was computed as the average value over configurations generated during one training step. $n$ is a normalizing factor that corresponds to the size of the system, which in this case is equal to the number of atoms in the 2nn sphere of influence (18). Each step of the training consisted of generating 50 2nn atomic configurations
    % \protect\footnotemark
    , with BSD3 type of defect. 
    Three different values of $\epsilon$ were added to the $\bm{\phi}$ parameters of $R_{\bm{\phi}}$ distribution from Eq.~(\ref{eq:lnZ_fq_Jensen_loss}) as explained by Eq.~(\ref{eq:phi_constant}).}
    \label{fig:training_w}
\end{figure}

% \footnotetext{Different sizes of training batches were tested, and a batch size of 50 was sufficient to obtain good training performance for $|\bm{y}| = 8$. In the case of larger $\bm{y}$ the size of training batches should also be increased.}

When the model is randomly initialized, meaning there's an equal chance of having Pu or U at each atomic site, it will naturally generate configurations with around 50\% Pu content. However, if a different Pu concentration is desired in the generated configurations, the model needs to adjust $\bm{\phi}$ to increase or decrease the local probability of Pu, which can extend the training time. We can mitigate this by altering the initialization of the model parameters. The simplest approach is to add a constant to the predicted $\bm{\phi}$ parameters:

\begin{equation}\label{eq:phi_constant}
    \bm{\phi}_{\epsilon} = \left( \phi^{(1)} + \epsilon, \phi^{(2)} + \epsilon, ..., \phi^{(n)} + \epsilon \right),
\end{equation}

\noindent
where $n$ is the size of the vector $\bm{\phi}_{\epsilon}$. We can try to find a suitable $\epsilon$ numerically, but a good candidate for $\epsilon$ can be also computed analytically using Pu concentration $y_{\mathrm{Pu}}$. Values $\bm{\phi} \in \mathbb{R}$ can vary from $-\infty$ to $+\infty$, while $y_{\mathrm{Pu}}$ varies from 0 to 1, therefore, to compute $\epsilon$ we need to map $y_{\mathrm{Pu}}$ into the domain of $\bm{\phi}$. It can be done by applying logit function to $y_{\mathrm{Pu}}$, so we obtain:

\begin{equation}\label{eq:logit_y}
    \epsilon = \ln{\frac{y_{\mathrm{Pu}}}{1 - y_{\mathrm{Pu}}}}.  
\end{equation}

An interesting property of Eq.~(\ref{eq:logit_y}) is that when $y_{\mathrm{Pu}} = 0.5$, $\epsilon$ is also equal to 0 and introduces no change to $\bm{\phi}_\epsilon$. This makes sense as if our target is to generate configurations with 50\% Pu, equiprobable sampling is enough to obtain the required concentrations.

We can evaluate the impact of $\epsilon$ defined in Eq.~(\ref{eq:logit_y}) on enhancing model training performance. This is achieved by monitoring the variation of the logarithm of the configuration probability $\ln{\left(w'(\bm{x}_\mathrm{c})\right)}$ within $f_T(\bm{x}_\mathrm{c})$ of the loss function described in Eq.~(\ref{eq:lnZ_fq_Jensen_loss}), as illustrated in Fig.~\ref{fig:training_w} for MOX with 10 \% Pu. The objective is to converge this component as quickly as possible.

We can see that choosing a custom starting point to initialize the model parameters can significantly influence the time needed for the model to converge. From now on in the next experiments, we will always initialize $\bm{\phi}$ parameters of $R_{\bm{\phi}}$, using Eq.~(\ref{eq:phi_constant}) and Eq.~(\ref{eq:logit_y})  formula. However, it is important to note that the choice of constant $\epsilon$ as expressed in Eq.~(\ref{eq:phi_constant}), while optimal in our case, might not be as optimal when applied to other parameters or different systems, where there might be other factors that influence the final $R_{\bm{\phi}}$.

\subsection{Benchmarking the IVAE approach in the MOX system}\label{sec:Accuracy_of_IVAE_approach}

In this section, we test the accuracy of IVAE predictions. Calculating the exact partition function in a MOX environment is challenging, but it becomes comparatively easier in smaller environments due to the reduced extent of the configuration space involved. To assess the accuracy of the IVAE models, we compare their partition function predictions with numerically computed values for 1nn and 2nn spheres of influence from preexisting databases. It is worth reminding, however, that using only 1nn or 2nn atoms provides a weak approximation of defect concentration. This approach is used solely to evaluate the performance of the IVAE models.

To achieve this we will use two databases. The first one is the database of 1nn BSD3 defect formation energies from \cite{karcz2023semi}. The second one was prepared during the work on the IVAE method and consists of 262144 configurations of 2nn BSD3 defect formation energies ($2^{18}$, where 18 is the number of atoms in the 2nn sphere). 
% In contrast to previous databases listed in Table~\ref{tab:databases}, this one features randomly filled external environments around the 2nn sphere for each configuration. This was done in a manner that ensures the total targeted Pu concentration remains constant, regardless of the local atomic composition of the 2nn sphere. While we could reduce the number of configurations in this database by focusing solely on the sphere's symmetries and ignoring the external environment, we've chosen to retain them. Since the external environment is random, even symmetric configurations will yield different formation energies. Therefore we prepared a larger 2nn database to perform a more precise computation of the partition function, that takes into account the variations of the external environment. 

\begin{figure*}[p]
    \vspace{0.5cm}
    \centering
    {\includegraphics[width=0.4\textwidth]{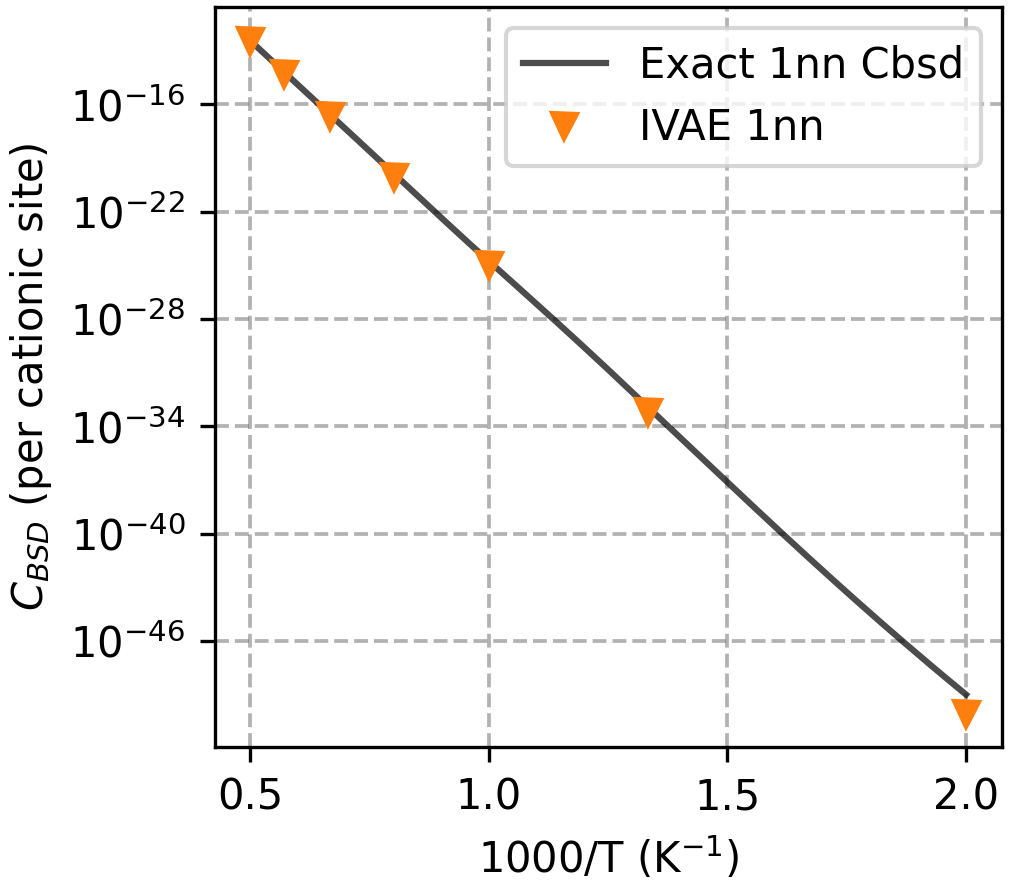}}
    \hspace{30pt}
    {\includegraphics[width=0.4\textwidth]{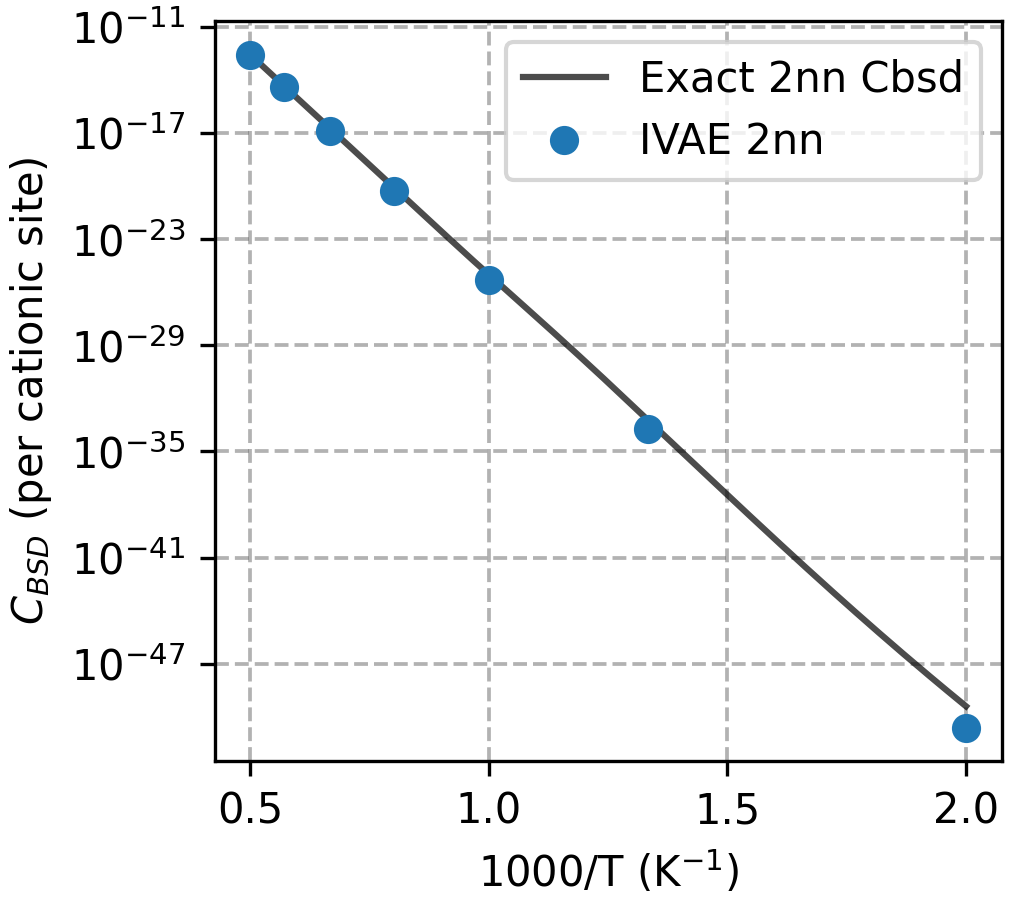}}
    % First goest to the list of figures. Second is the description underneath
    \caption[IVAE BSD3 defect concentration for 1nn and 2nn sphere of influence in MOX 10\% Pu]{IVAE BSD3 defect concentration for 1nn and 2nn sphere of influence in MOX 10\% Pu. Exact computations were performed using the 1nn BSD3 database from \cite{karcz2023semi} and the 2nn BSD3 database developed in this study. The computations utilized the formula given by Eq.~(\ref{eq:CdT_concentration_compressed}), with $y_{\mathrm{Pu}} = 0.1$ from Eq.~(\ref{eq:p_prime_concentration}). Predictions were generated by models achieving a minimum accuracy of 96.5\%, determined by comparing predicted defect concentrations to analytically computed values. For both 1nn and 2nn systems, $|\bm{y}| = 8$ was used. The parameters $\bm{\phi}$ of the predicted $R_{\bm{\phi}}$ distribution were initialized using the formula provided by Eq.~(\ref{eq:logit_y}).}
    \label{fig:IVAE_acc_10Pu}
\end{figure*}

\begin{figure*}[p]
    \vspace{0.5cm}
    \centering
    {\includegraphics[width=0.47\textwidth]{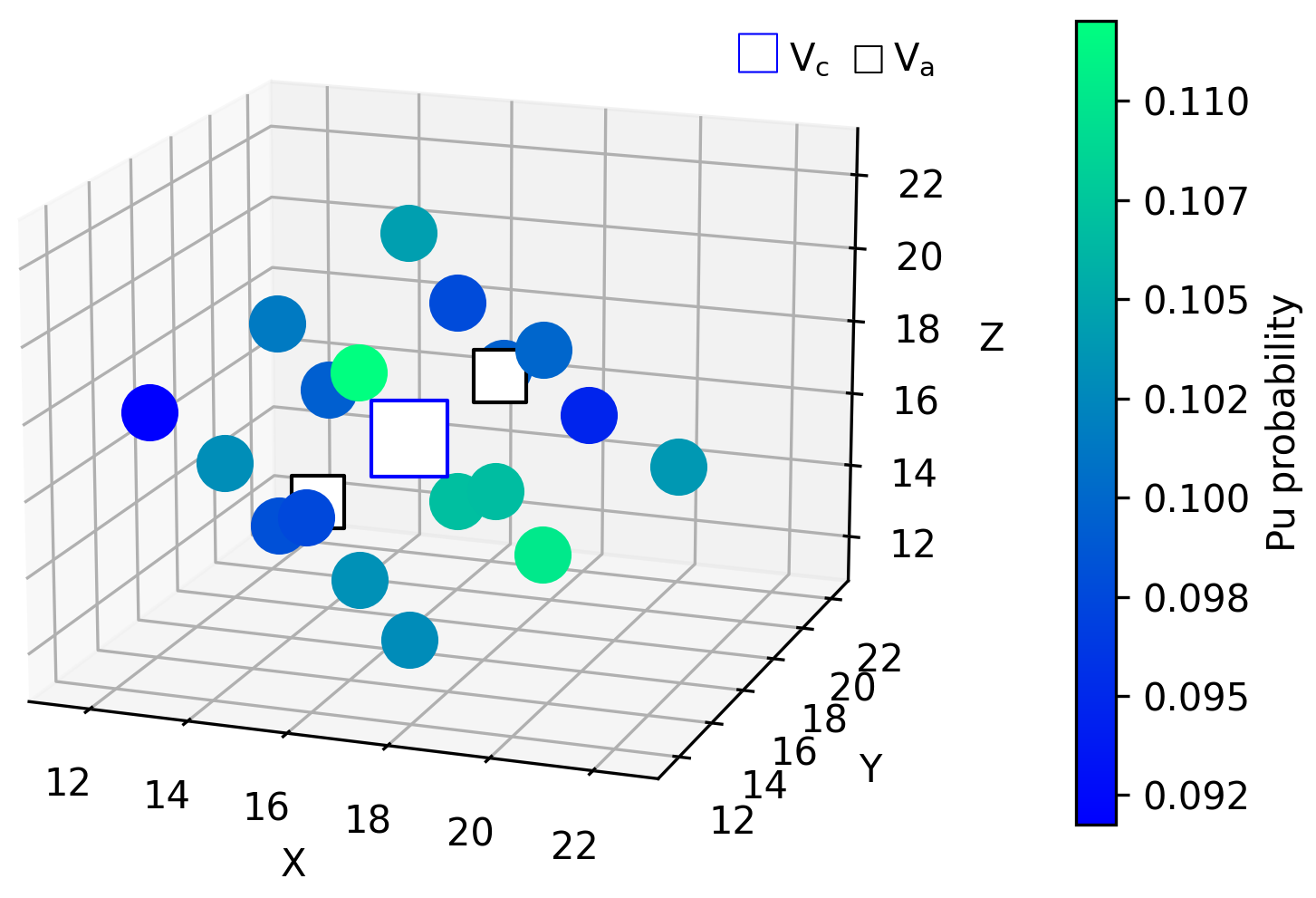}}
    \hspace{30pt}
    {\includegraphics[width=0.33\textwidth]{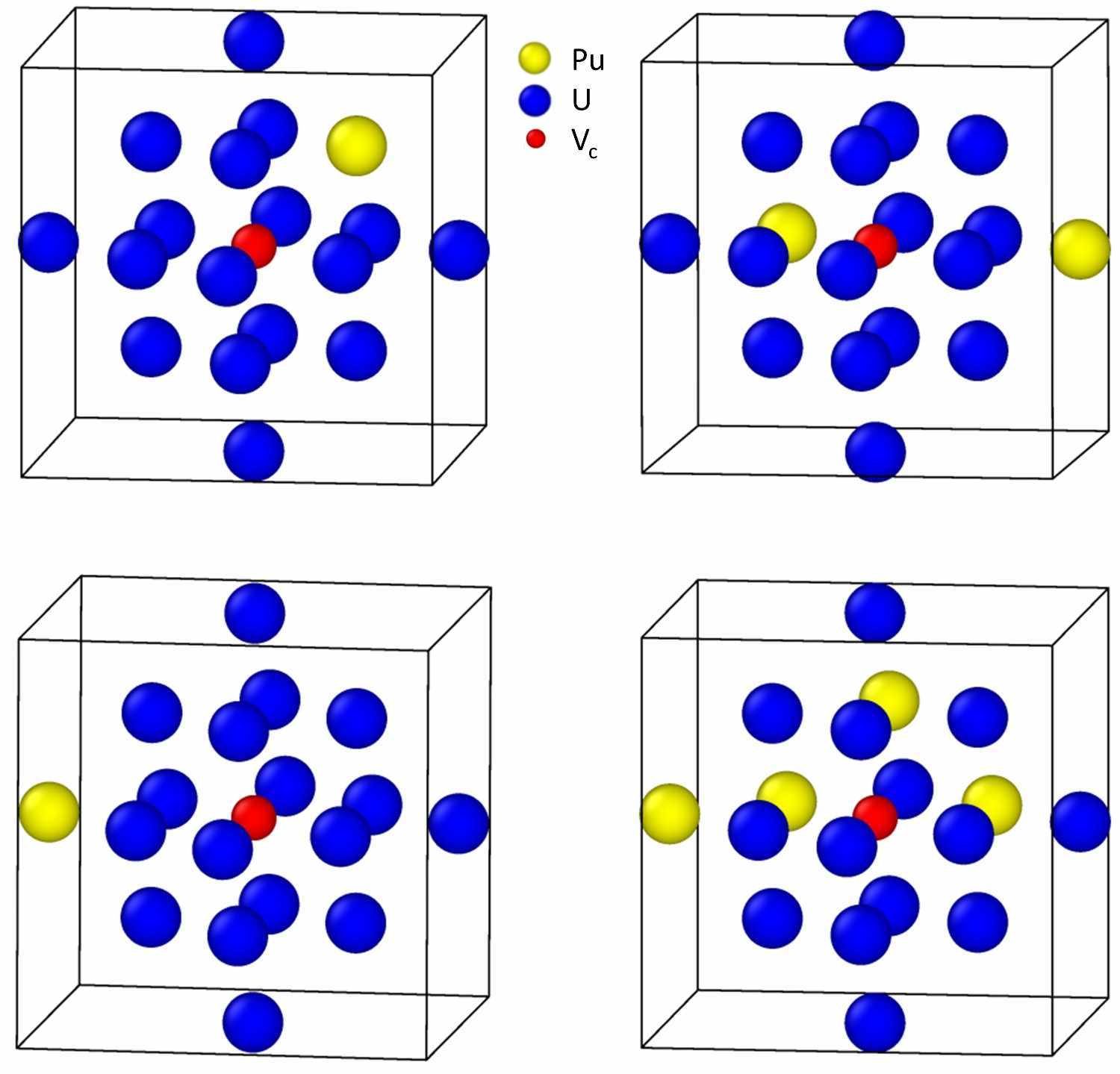}}
    % First goest to the list of figures. Second is the description underneath
    \caption[3D visualization of the IVAE probability distribution for 2nn atomic configurations for BSD3 in MOX, T = 500 K, with 10\% global Pu concentration]{On the left is a 3D representation of the IVAE probability distribution, which represents the average occupancy on each atomic site surrounding the BSD3 defect for 2nn atomic configurations in MOX at T = 500 K, with a 10\% global Pu concentration. On the right, a few samples of 2nn configurations sampled from the multidimensional distribution $R_{\bm{\phi}}$. The 3D visualization of the IVAE probability distribution was computed by averaging probabilities for each cationic site predicted by $R_{\bm{\phi}}$. 
    %The precise sampling process to obtain the spheres on the right involves first sampling $\bm{y}$ from $P(\bm{y})$, and then $\bm{x}_{\mathrm{c}}$ from $R_{\bm{\phi}} (\bm{x}_{\mathrm{c}} | \bm{y})$, as explained in Sec.~\ref{sec:Adding_interactions}.
    }
    \label{fig:IVAE_3D_10Pu}
\end{figure*}

\begin{figure*}[t]
    \vspace{0.5cm}
    \centering
    {\includegraphics[width=0.4\textwidth]{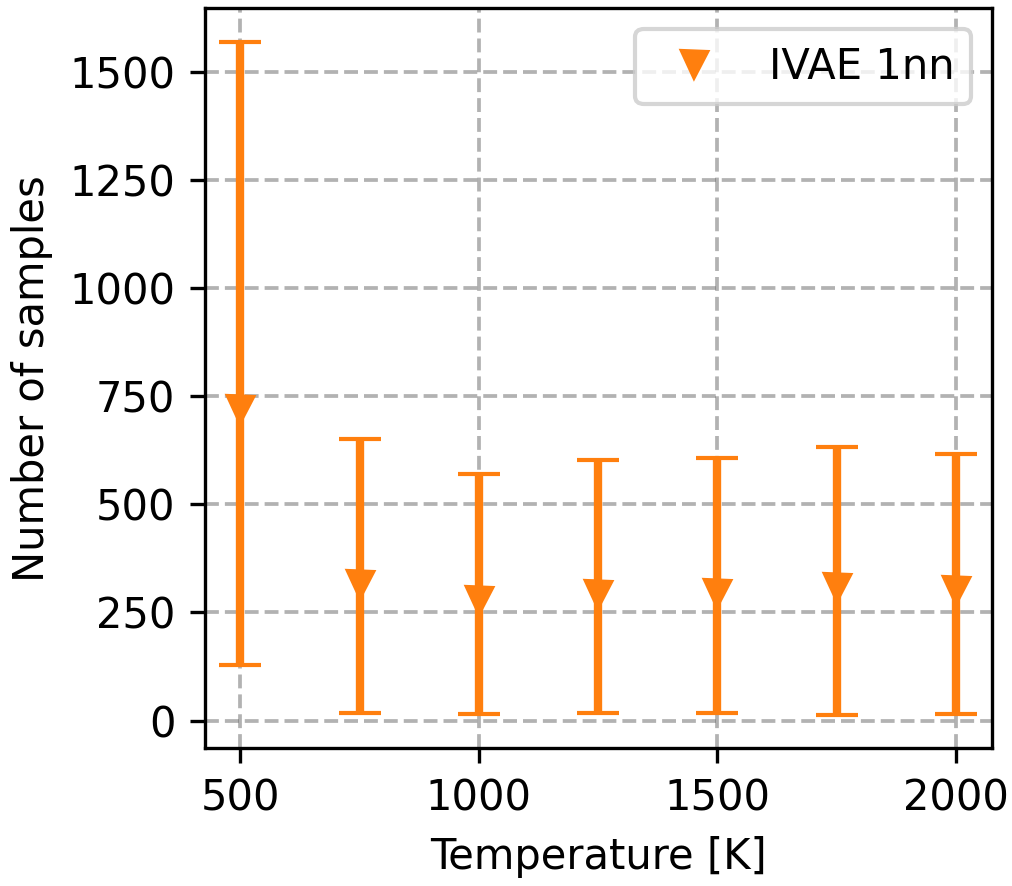}}
    \hspace{30pt}
    {\includegraphics[width=0.4\textwidth]{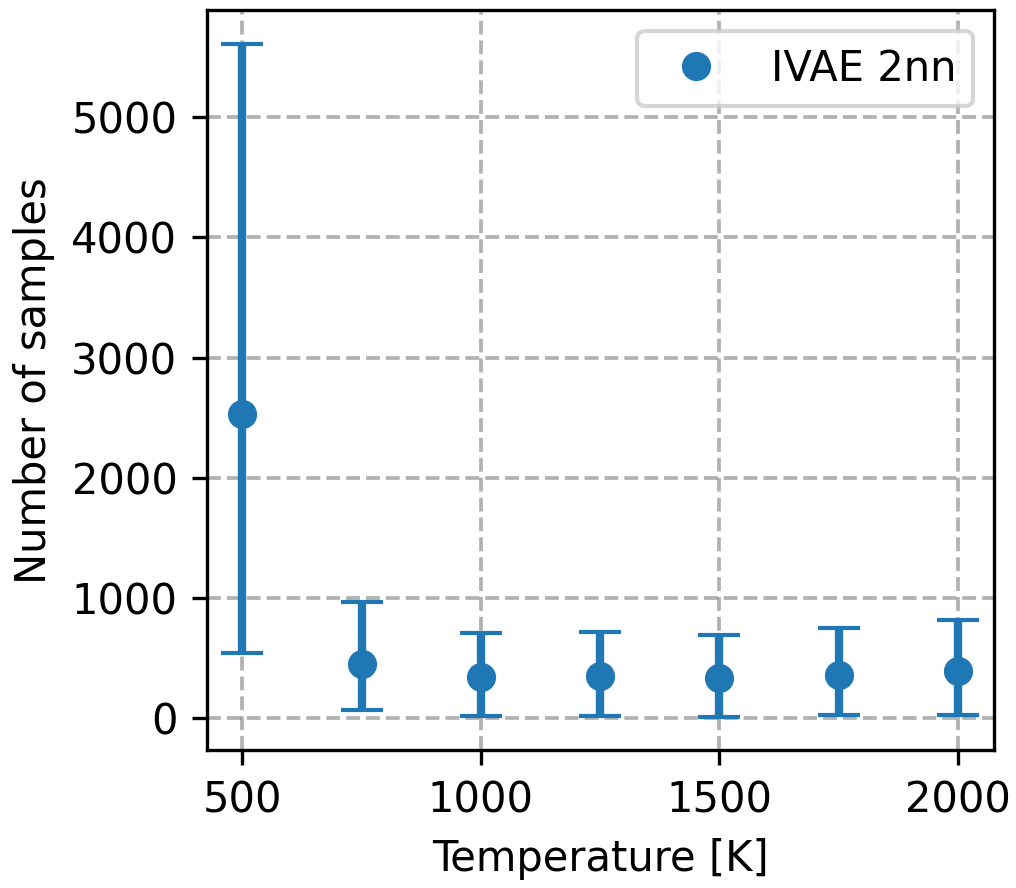}}
    % First goest to the list of figures. Second is the description underneath
    \caption[The number of samples required to achieve 96.5\% accuracy in defect concentration by IVAE models in 1nn and 2nn systems.]{The number of samples required to achieve 96.5\% accuracy in defect concentration by IVAE models in 1nn and 2nn systems. Experiment done for MOX with 50\% global Pu concentration. Each estimation of the training cost is shown as a 95\% confidence interval of the predictions of 40 IVAE models. The number of samples specifies explicitly the amount of atomic configurations, and thus the required energy minimizations with the CRG potential \cite{cooper2014many}, that the model needed upon training to generate to achieve at least 96.5\% accuracy of the defect concentration.}
    %in Fig.~\ref{fig:IVAE_acc_50Pu}
    \label{fig:IVAE_train_time_50Pu}
\end{figure*}

We have tested IVAE model predictions for two sizes of the local environment and for three different global Pu concentrations: 10\%, 50\%, and 90\%. The results for 10\% Pu concentration can be seen in Fig.~\ref{fig:IVAE_acc_10Pu} and are compared to the exact computations of defect concentration. 
%The exact computations of the defect concentration in the previously described 1nn and 2nn sphere of influence databases were performed using Eq.~(\ref{eq:CdT_concentration_compressed}) formula, where $y_{\mathrm{Pu}}$ from Eq.~(\ref{eq:p_prime_concentration}) was set to 0.1, 0.5 and 0.9 to obtain the appropriate concentrations. 

The results in Fig.~\ref{fig:IVAE_acc_10Pu} represent the predictions made by IVAE models, which were trained to attain a minimum accuracy of 96.5\% measured by comparing the predicted defect concentration values to analytically computed ones. The amount of sampled atomic configurations required by the models to achieve the specified above accuracy threshold for MOX with 50\% Pu is shown in Fig.~\ref{fig:IVAE_train_time_50Pu}. The results for other concentrations were similar.

We have used a simple neural network architecture with one hidden layer of 1024 neurons and the SELU activation function, for both encoder $R_{\bm{\phi}}$ and decoder $Q_{\bm{\theta}}$ parts of the IVAE model. The temperature parameter of the Gumbel softmax function was set to $\tau = 1/10$ and all experiments used Adam optimizer with TensorFlow 2.8 default parameters. We used a Bernoulli distribution as $P(\bm{y})$. We tested various values of $|\bm{y}|$ and training batch sizes, and we selected $|\bm{y}| = 8$ with training batch sizes of 50 atomic configurations per training step, as this configuration yielded satisfactory results.

In Fig.~\ref{fig:IVAE_acc_10Pu} IVAE demonstrates its capability to compute defect concentrations with high accuracy, as illustrated. Moreover, as depicted in Fig.~\ref{fig:IVAE_train_time_50Pu}, this is achieved without significant computational cost where the total number of configurations in the 1nn and 2nn databases are 4096 and 262144, respectively. The cost increases notably only at low temperatures, where fewer critical configurations contribute significantly to the defect concentration calculation, posing challenges in their identification. Additionally, achieving accuracies exceeding 96.5\% necessitates increased total training time for obtaining precise defect concentrations.

Additionally, the multidimensional probability distribution of configurations $R_{\bm{\phi}} (\bm{x}_{\mathrm{c}} | \bm{y})$ can be reduced and visualized in three-dimensional space, as shown in Fig.~\ref{fig:IVAE_3D_10Pu}. This representation averages the probabilities of all components of $R_{\bm{\phi}} (\bm{x}_{\mathrm{c}} | \bm{y})$ for each atomic site from 50 generated configurations, for which the calculated partition function was highest.

As we can see, the model behaves as expected, correctly generating configurations with occupancy probabilities for each site that oscillates around the global Pu concentration. One could attempt to interpret the differences between individual sites as an influence of possible local order. However, it is important to note that this type of visualization of a reduced probability distribution might not accurately capture all atomic interactions in the data.

A good example can be given in the case of the Ising system at low temperatures. At low temperatures, all of the spins in an Ising system tend to align in one direction — either all spins up or all spins down. If we proceed as in, for example, Fig.~\ref{fig:IVAE_3D_10Pu}, using IVAE to learn the partition function of the Ising system in such a scenario, and then take the last 50 configurations of spins, which are either all up or all down, and compute the accumulated probability at each spin site, we would find that each individual spin site has a 50\% chance to be either up or down.  This clearly shows that to capture the all-spins-up or all-spins-down property, a more sophisticated analysis is required. One way to make this type of analysis much more interesting is to inspect the probability distribution of descriptors instead, which can be explored in future work.

Nevertheless, with easy access to all components of $R_{\bm{\phi}} (\bm{x}_{\mathrm{c}} | \bm{y})$ and other elements of the model architecture, IVAE provides a flexible sampling framework, where the sampling process can be monitored in high detail.

So far, we have tested the IVAE models, evaluating their accuracy and training performance across different systems and settings. We have shown the model predictions in MOX for different global Pu concentrations and we also proposed a way to initialize the model parameters to minimize the training cost. With the model well-tested, in the following part, we will apply the IVAE to larger-scale experiments in $\mathrm{(U, Pu)O_2}$ system.

\subsection{Measuring the range of influence in MOX with IVAE}\label{sec:Measureing_range_of_influence_IVAE}

\begin{figure*}[h!tbp]
    \vspace{0.5cm}
    \centering
    {\includegraphics[width=0.4\textwidth]{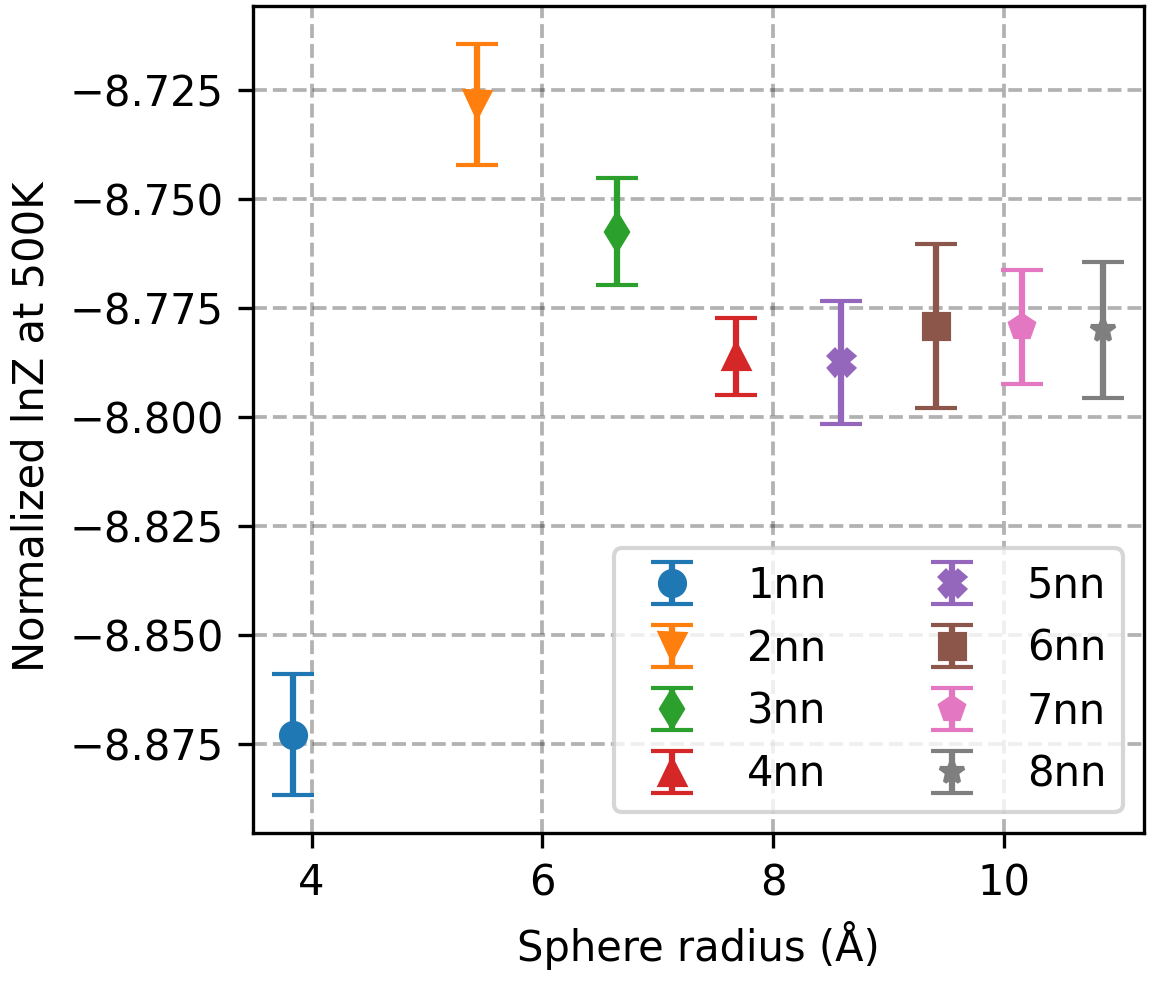}}
    \hspace{30pt}
    {\includegraphics[width=0.4\textwidth]{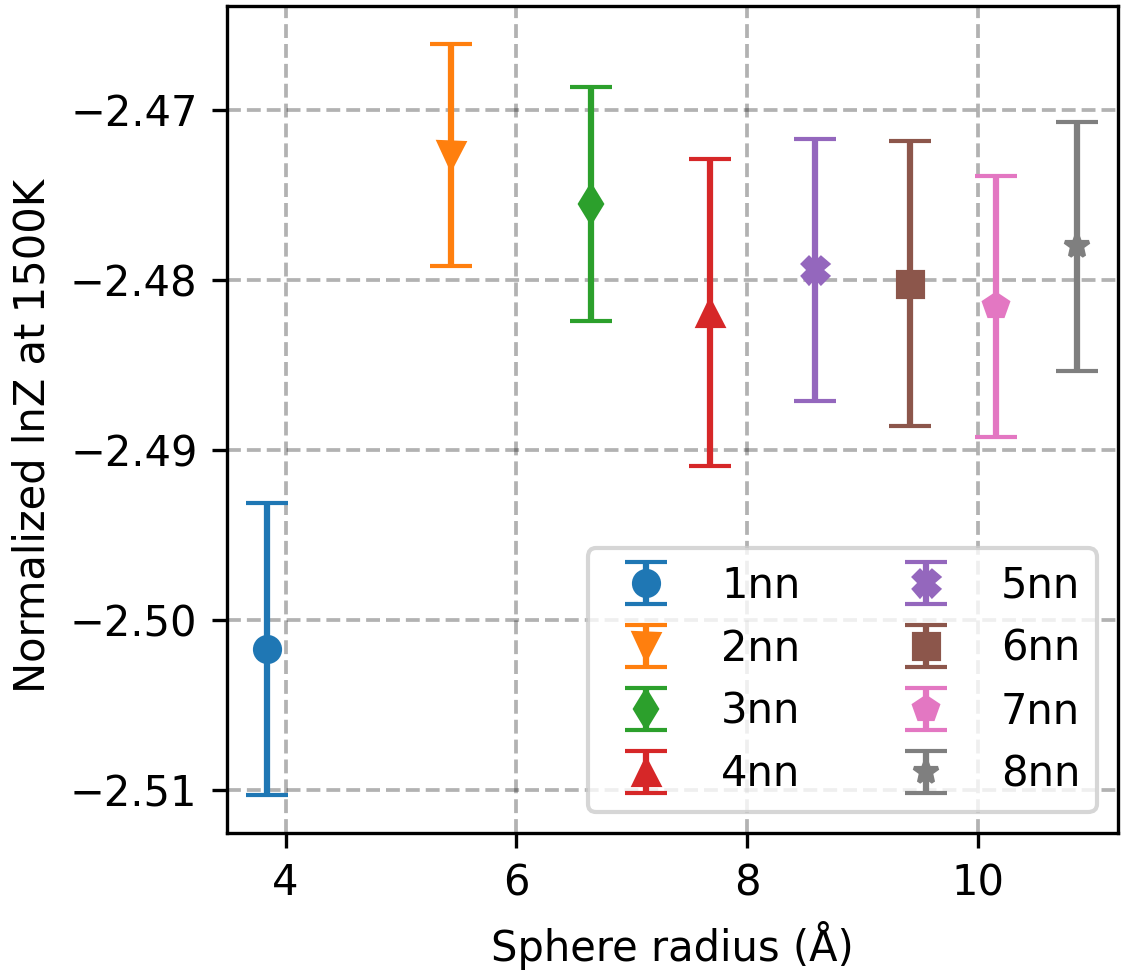}}
    % First goest to the list of figures. Second is the description underneath
    \caption[Measuring range of influence with IVAE]{Measuring range of influence with IVAE. Both graphs show the IVAE predictions in MOX with 50\% global Pu concentration for two different temperatures: 500 K on the left and 1500 K on the right. IVAE was used to predict the normalized partition function, which corresponds to the estimation of $\ln{(Z'_T)}/n$, where $n = 12$ and $Z'_T$ comes from Eq.~(\ref{eq:_Zprime_N}). Error is measured as a 95\% confidence interval of 40 estimations of the partition function.}
    \label{fig:IVAE_range_T500_T1500}
\end{figure*}

In this section, we use the IVAE model to measure the range of influence of local atomic environments. We apply IVAE to systems of progressively increasing size, ranging from 1nn to 8nn spheres of influence around the defect. We employ a constant reference supercell with a 50\% Pu concentration, where atoms in the center are replaced with pre-prepared 1-8nn supercells. Since we are working with MOX with 50\% Pu concentration, we can also simplify the mathematical formulas for the computation of the defect concentration. For 50\% Pu concentration, $w'(\bm{x}_\mathrm{c})$ is constant and equal $\frac{1}{N}$, where $N$ is the total amount of configurations in the configuration space. Instead of including $w'(\bm{x}_\mathrm{c})$ within $f_T(\bm{x}_\mathrm{c})$, we can write:

\begin{equation}\label{eq:CdT_concentration_compressed_Zf_N}
         C_{\mathrm{BSD}}(T) = \frac{Z'_T}{N},
\end{equation}

\noindent
where:

\begin{equation}\label{eq:_Zprime_N}
         Z'_T = \sum_{\bm{x}_\mathrm{c} \in \bm{X}} \exp\left({-\frac{E_{\mathrm{BSD}}^{\mathrm{f}}(\bm{x}_\mathrm{c})}{k_{\mathrm{B}}T}} \right).
\end{equation}

Therefore, for the sake of this experiment, we use IVAE to estimate $Z'_T$. IVAE is set up to generate 1nn spheres of influence, while the U/Pu distribution for other atoms in the local environment (from 2nn to 8nn) is determined randomly. Since we are generating 1nn spheres only, comprising 12 atoms, the $\bm{\phi}$ vector in the IVAE architecture has 12 elements. This approach allows the calculation of formation energy to be primarily influenced by 1nn atoms, with the noise introduced by 2-8nn atoms. We measure the range of influence of local atomic environments for two different temperatures: 500 K and 1500 K. The results are shown in Fig.~\ref{fig:IVAE_range_T500_T1500}.

Similarly as in Sec.~\ref{sec:Accuracy_of_IVAE_approach} we have used a simple neural network architecture with one hidden layer of 1024 neurons and the SELU activation function, for both encoder $R_{\bm{\phi}}$ and decoder $Q_{\bm{\theta}}$ parts of the IVAE model. The temperature parameter of the Gumbel softmax function was set with $\tau = 1/10$ and all experiments used Adam optimizer with TensorFlow 2.8 default parameters. We used a Bernoulli distribution as $P(\bm{y})$, with the size of the input vector  $|\bm{y}| = 8$. Models were trained iteratively, generating 50 atomic configurations in each training step.

\begin{figure*}[h!tbp]
    \vspace{0.5cm}
    \centering
    {\includegraphics[width=0.35\textwidth]{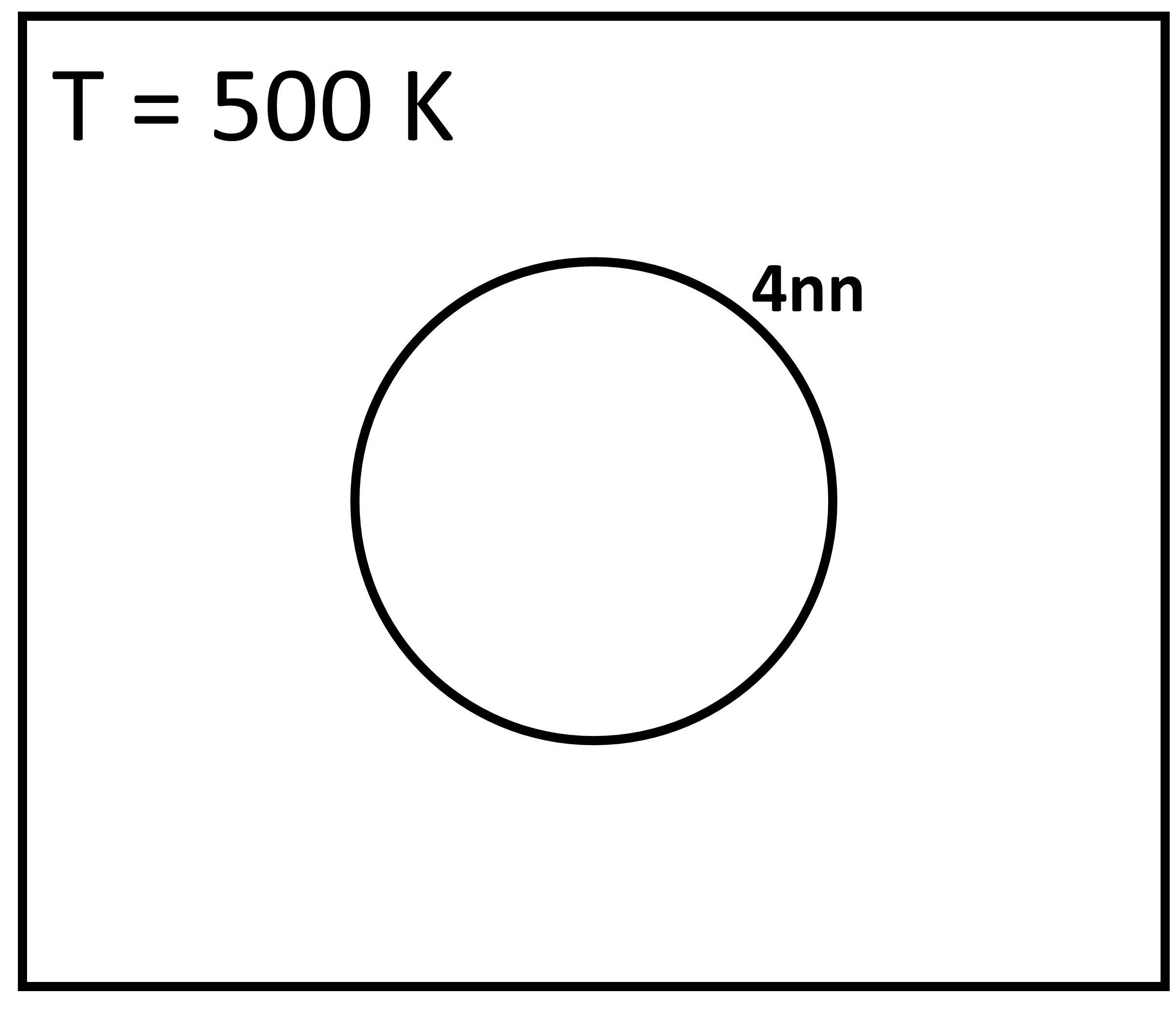}}
    \hspace{50pt}
    {\includegraphics[width=0.35\textwidth]{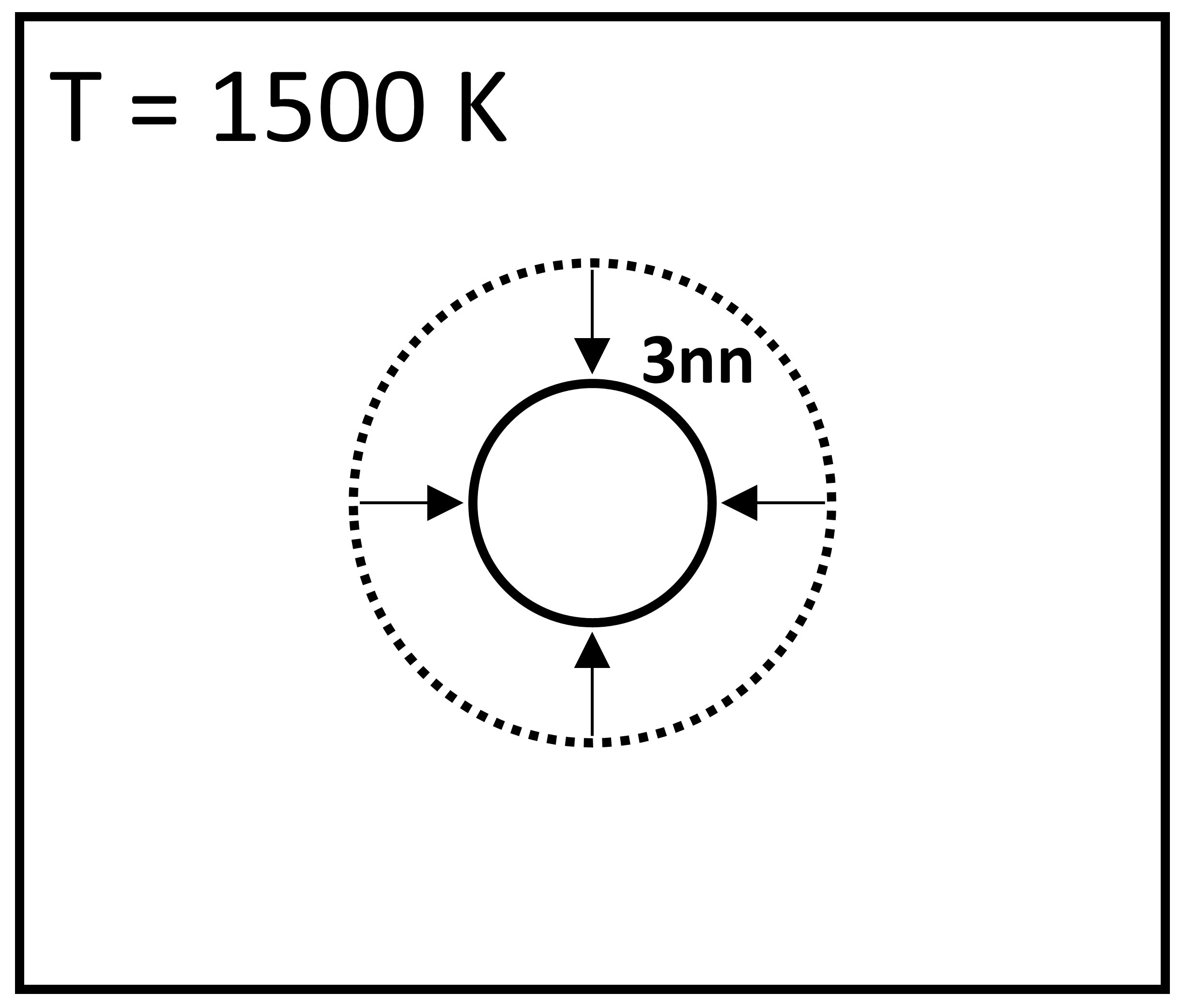}}
    % First goest to the list of figures. Second is the description underneath
    \caption[Graphical visualization of how temperature affects the range of influence]{Graphical visualization of how temperature affects the range of influence. Graphs show a two-dimensional representation of the 6x6x6 MOX supercell, with the range of influence marked in the middle for two different temperatures: 500 K and 1500 K.}
    \label{fig:IVAE_range_T500_T1500_diagram}
\end{figure*}

\begin{figure}[h!t]
    \vspace{0.5cm}
    \centering
    \includegraphics[width=0.48\textwidth]{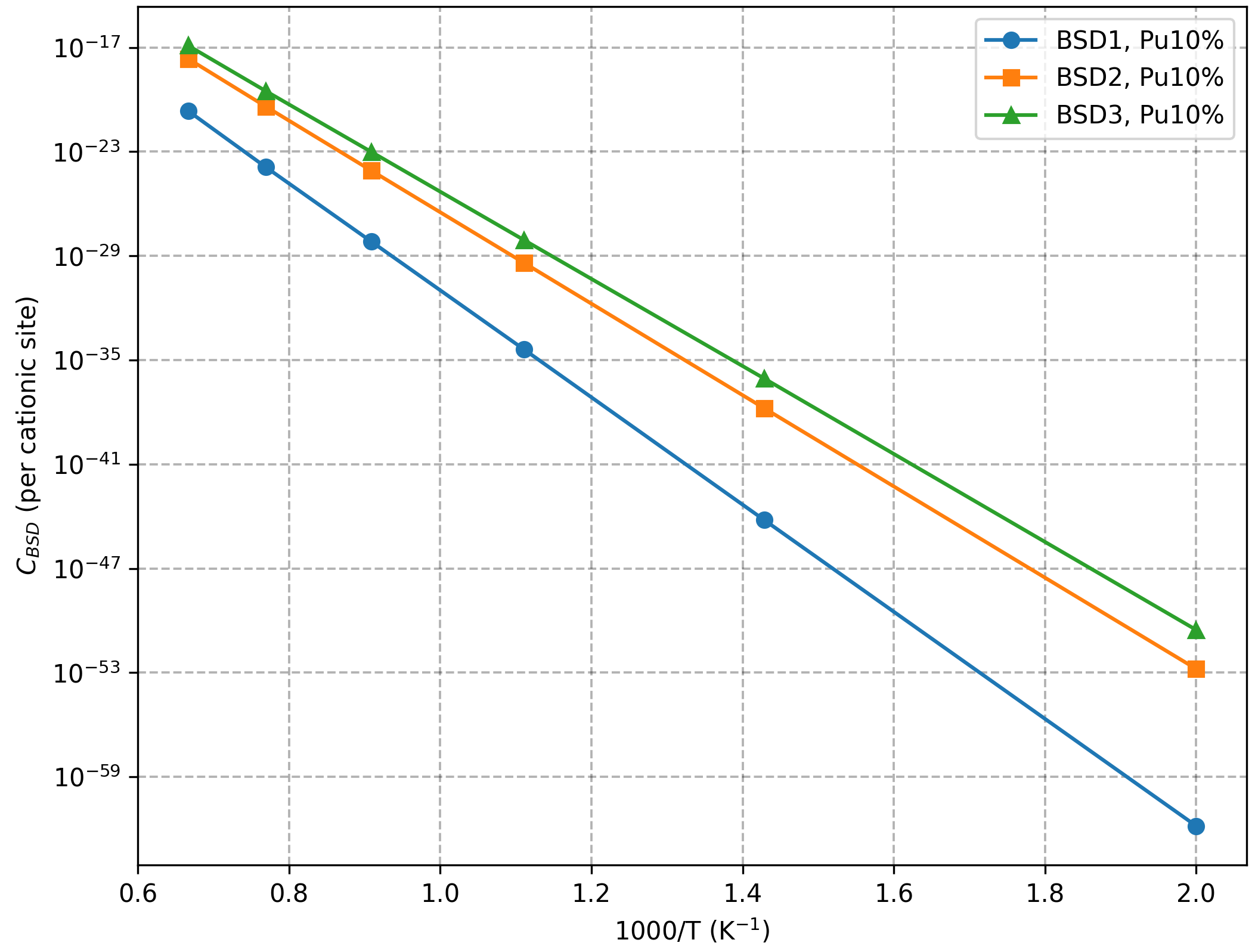}
    % First goest to the list of figures. Second is the description underneath
    \caption[Comparison of the defect concentration $C_{\mathrm{BSD}}(T)$ of different BSD defects in $\mathrm{(U, Pu)O_2}$ with 10\% Pu computed from 4nn IVAE predictions]{Comparison of the defect concentration $C_{\mathrm{BSD}}(T)$ of different BSD defects in $\mathrm{(U, Pu)O_2}$ with 10\% Pu computed from 4nn IVAE predictions. Each point represents the mean of the predictions of IVAE models. The details about the predictions and their variance are summarized in Table~\ref{tab:IVAE_cbsd_table}.}
    \label{fig:IVAE_cbsd_Pu10}
\end{figure}

\begin{figure}[h!t]
    \vspace{0.5cm}
    \centering
    \includegraphics[width=0.48\textwidth]{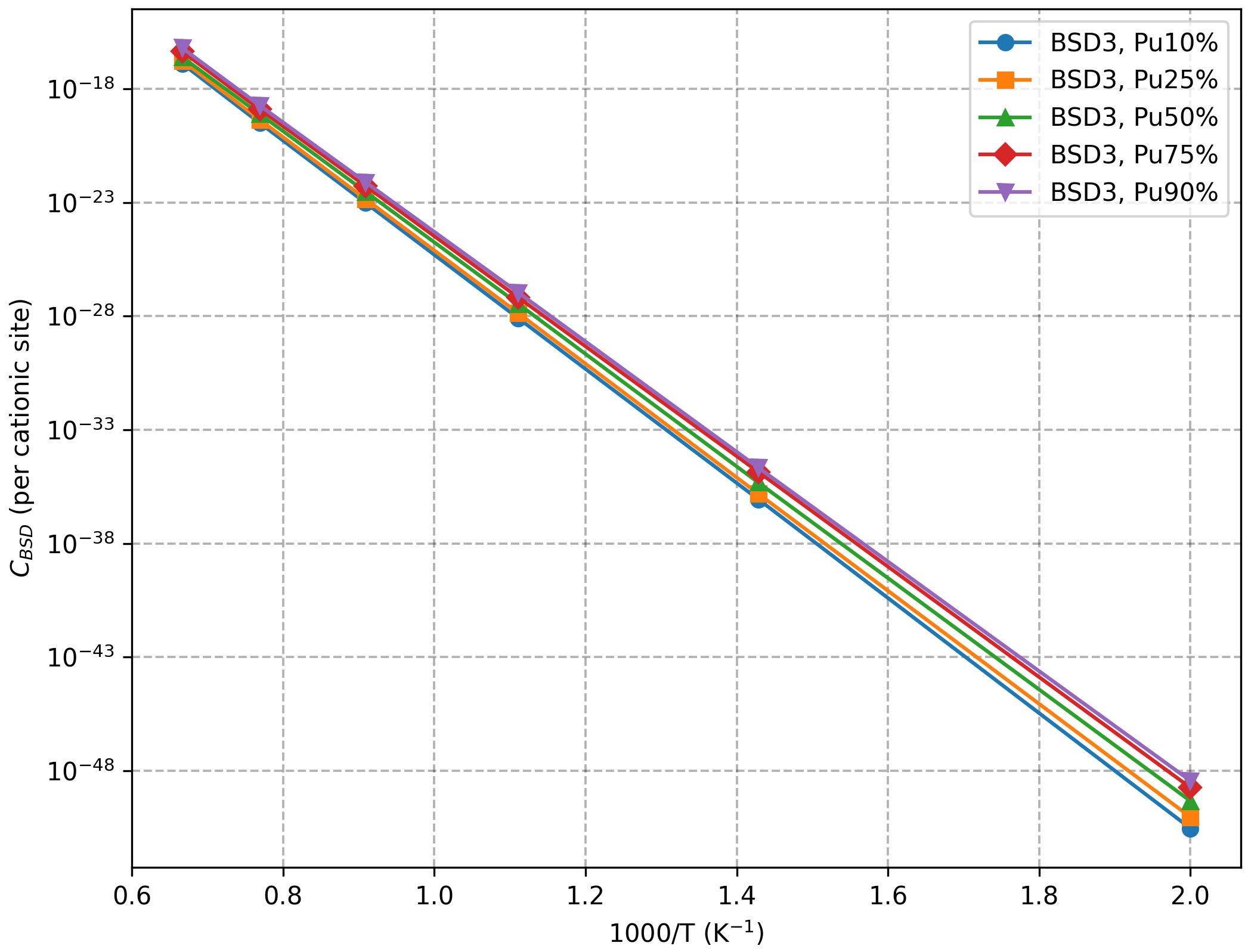}
    % First goest to the list of figures. Second is the description underneath
    \caption[Comparison of the defect concentration $C_{\mathrm{BSD}}(T)$ for BSD3 at different Pu concentrations computed from 4nn IVAE predictions]{Comparison of the defect concentration $C_{\mathrm{BSD}}(T)$ for BSD3 at different Pu concentrations computed from 4nn IVAE predictions. Each point represents the mean of the predictions of IVAE models. The details about the predictions and their variance are summarized in Table~\ref{tab:IVAE_cbsd_table}.}
    \label{fig:IVAE_cbsd3_diffPu}
\end{figure}

\begin{figure}[h!t]
    \vspace{0.5cm}
    \centering
    \includegraphics[width=0.48\textwidth]{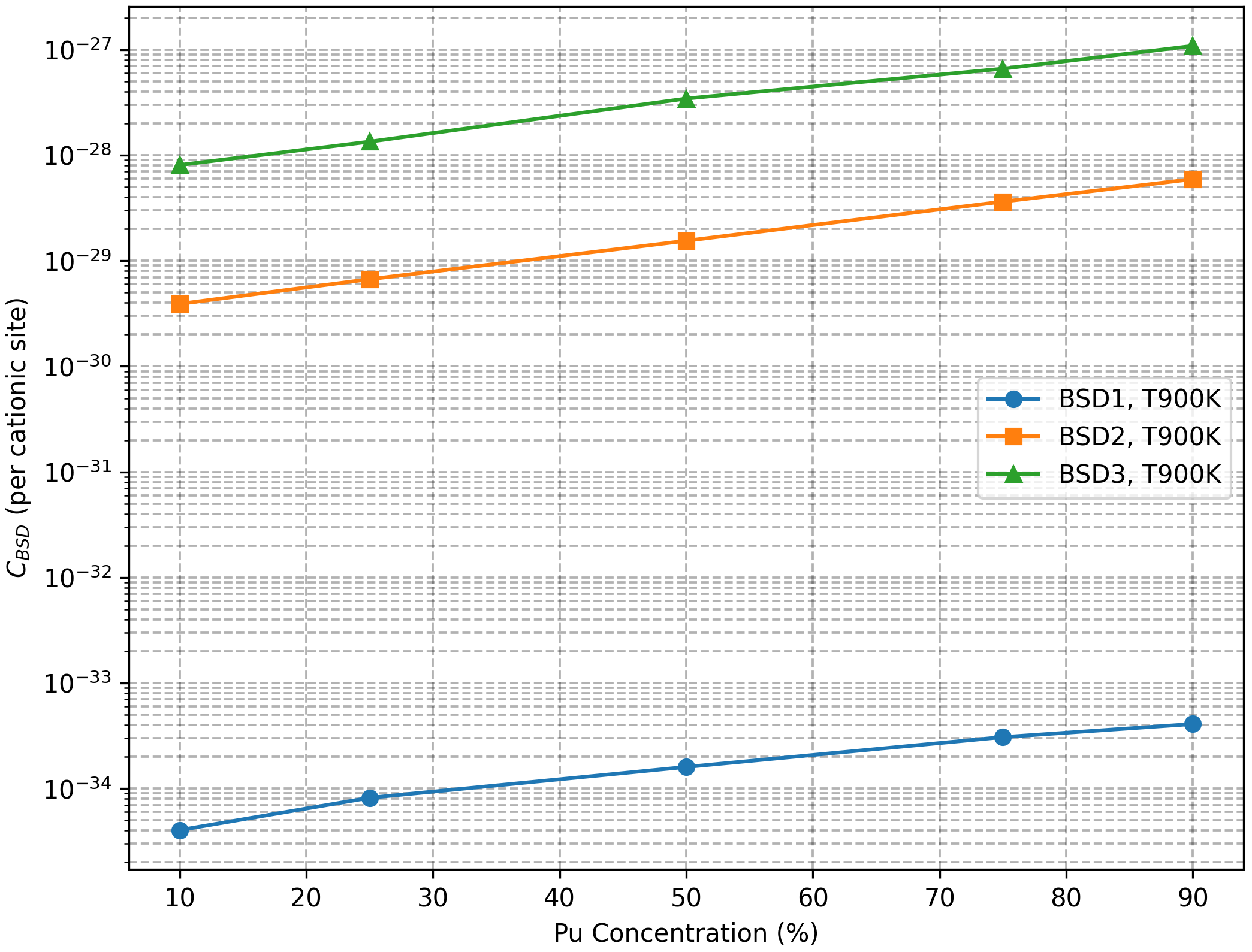}
    % First goest to the list of figures. Second is the description underneath
    \caption[Comparison of the defect concentration $C_{\mathrm{BSD}}(T)$ for different BSD types at various Pu concentrations in MOX at 900 K, computed from 4nn IVAE predictions]{Comparison of the defect concentration $C_{\mathrm{BSD}}(T)$ for different BSD types at various Pu concentrations in MOX at 900 K, computed from 4nn IVAE predictions. Each point represents the mean of the predictions of IVAE models. The details about the predictions and their variance are summarized in Table~\ref{tab:IVAE_cbsd_table}.}
    \label{fig:IVAE_cbsd_T900K}
\end{figure}

One advantage of IVAE over the previous MDN approach from \cite{karcz2023semi} is that it includes the influence of temperature in its predictions (reflected by the $\frac{1}{k_{\mathrm{B}}T}$ factor, as we can see, for example, in Eq.~(\ref{eq:CdT_concentration_compressed}), enabling a deeper understanding of the physical behavior of our system \footnote{Currently, the model does not account for the entropic effects of finite temperature on the calculated formation energies. However, this limitation is due to the atomic-scale calculations used to compute the formation energy, and the IVAE model can also be adapted to incorporate this factor.}. For instance, we can observe changes in the range of influence when adjusting the temperature. At lower temperatures, the models' predictions converge around a 4nn sphere of influence. However, as the temperature increases, this convergence occurs sooner, closer to the 3nn sphere, as illustrated in Fig.~\ref{fig:IVAE_range_T500_T1500_diagram}. The calculation of defect concentration depends on the formation energy of atomic configurations, a local property primarily influenced by the atoms closest to the defect. The range of influence defines the limit beyond which atomic composition starts to have minimal impact on the formation energy computations. As shown in Fig.~\ref{fig:IVAE_range_T500_T1500_diagram}, this range of influence varies with temperature. This is because each shell's average contribution to formation energy decreases as one moves outward from the defect. At lower temperatures, the energy differences between shells are accentuated due to the Boltzmann factor $\frac{1}{k_{\mathrm{B}}T}$. This means that the energy gaps, such as between 3nn and 4nn, have a more pronounced effect at lower temperatures compared to higher temperatures. Consequently, achieving accurate predictions of partition function requires extending the analysis to include more distant shells at lower temperatures to ensure convergence. 

This result differs from findings in MDN work \cite{karcz2023semi} and the study by Bathellier \textit{et~al.} \cite{didier2022disorder}, where the 3nn distance was identified as a threshold beyond which additional atoms have negligible influence on formation energy calculations. It is important to note, however, that these studies employed a different methodology. Unlike IVAE, which generates distinct configurations for specific temperatures, previous methods used a single set of configurations to encompass the entire configuration space across all temperatures. This approach could be seen as measuring an average range of influence rather than the detailed analysis provided by IVAE.

\subsection{Defect concentration for different Pu concentrations and BSD types}\label{sec:def_con_diff_Pu_IVAE}

So far, we have tested IVAE in MOX system, measured its accuracy, optimized it, and applied it to measure the range of influence of local atomic configurations. Now, we will apply it to a larger-scale calculation of defect concentration. We use IVAE to predict the defect concentration of three types of Schottky defects for five Pu concentrations: 10\%, 25\%, 50\%, 75\%, 90\%, and temperatures ranging from 500 K to 1500 K.

We will use all the experiences and optimizations we have found so far in the previous experiments. To ensure that the local environment around the defect is large enough, we will task IVAE to generate 4nn spheres - as discussed in Sec.~\ref{sec:Measureing_range_of_influence_IVAE} and shown in Fig.~\ref{fig:IVAE_range_T500_T1500}. To assure good accuracy of the model's predictions, as discussed in Sec.~\ref{sec:Accuracy_of_IVAE_approach} and shown in Fig.~\ref{fig:IVAE_train_time_50Pu}, we train all IVAE models on 2000 atomic configurations. The only exception is the models for T = 500 K, where we used 6000 configurations instead. IVAE models were generating 50 configurations per training step, and we used a Bernoulli distribution as $P(\bm{y})$, with the size of the input vector $|\bm{y}| = 8$ (as discussed in Sec.~\ref{sec:Accuracy_of_IVAE_approach}). As before, we used a simple neural network architecture with one hidden layer of 1024 neurons and the SELU activation function for both the encoder $R_{\bm{\phi}}$ and decoder $Q_{\bm{\theta}}$ parts of the IVAE model. The temperature parameter of the Gumbel softmax function was set with $\tau = 1/10$, and all experiments used the Adam optimizer with TensorFlow 2.8 default parameters.

The environment surrounding the generated 4nn spheres is randomly filled with Pu and U to maintain the targeted global Pu concentration. The initial parameters of each IVAE model are initialized using $\epsilon$ from Eq.~(\ref{eq:logit_y}).

Figure~\ref{fig:IVAE_cbsd_Pu10} presents example predictions made by IVAE models for three types of defects at different temperatures and for 10\% Pu concentration. Figure~\ref{fig:IVAE_cbsd3_diffPu} compare different Pu concentrations for BSD3. The goal of IVAE is to predict the highest value of its loss function, which serves as a lower bound of the true partition function. Therefore, the values in the aforementioned figures correspond to the mean of the five highest predictions from each IVAE model. Complete results, including exact calculations and computed standard deviations, are summarized in Table~\ref{tab:IVAE_cbsd_table}. Additionally, we computed the effective formation energy of predicted defect concentrations using the following formula:

\begin{equation}
\begin{split}\label{eq:effEf}
        C_\mathrm{BSD}(T) &= \exp\left(-\frac{E_{\mathrm{eff}}^{\mathrm{f}}}{k_{\mathrm{B}}T}\right),\\
        E_{\mathrm{eff}}^{\mathrm{f}} &= -k_{\mathrm{B}}T\log{\left(C_\mathrm{BSD}(T)\right)}.
\end{split}
\end{equation}

\noindent
The computed effective energies are visualized in Table~\ref{tab:IVAE_eff_table}.

The variations observed in Figure~\ref{fig:IVAE_cbsd_Pu10}  stem from differences in BSD stability predicted by the CRG potential. These results align with previous findings in \cite{karcz2023semi}. Additionally, Figure~\ref{fig:IVAE_cbsd_T900K} demonstrates an increase in defect concentration with higher Pu concentrations, indicating a greater likelihood of defects forming in Pu-rich environments. This is consistent with previous CRG calculations, which show that the formation energy of defects is generally lower in $\mathrm{PuO_2}$ compared to $\mathrm{UO_2}$, making it energetically more favorable for defects to form as Pu concentration increases \cite{didier2022disorder}. However, the increase in defect concentration with higher Pu concentration, while noticeable for lower temperatures, quickly diminishes with the temperature increase, as shown in Figure~\ref{fig:IVAE_cbsd3_diffPu}.

% It is important to consider the differences between the MDN method from \cite{karcz2023semi} and IVAE when directly comparing their predicted defect concentrations. While both methods yield similar results, the methodologies for generating and utilizing configurations differ. MDN was trained using a prepared database where 1nn spheres were selected with varying Pu content (0-100\%), and the environment around these spheres (2nn to 4nn) was arranged randomly. During final supercell preparations for energy minimization, MDN utilized a constant reference MOX supercell where atoms at the center were replaced with those from the prepared spheres. This could result in the variations in the global Pu concentration. In contrast, IVAE employed a different approach. For the results in Table~\ref{tab:IVAE_cbsd_table}, IVAE was tasked with generating complete 4nn spheres of influence. To create full 6x6x6 supercells for energy minimization, the environment around these generated spheres was randomly filled with Pu while maintaining the targeted Pu concentration (Sec.~\ref{sec:IVAE_sphere_external_environment}). Furthermore, IVAE focused on specific values of $C_{\mathrm{BSD}}$ for each temperature, generating different sets of configurations for each scenario, unlike MDN which relied on a single pre-prepared dataset.

%Similarly to the case of MDN, 
It would be interesting to compare the BSD defect concentrations obtained with the approach discussed in this chapter with experimental measurements in $\mathrm{(U, Pu)O_2}$. However, to the authors' knowledge, no such study currently exists in the literature. From the perspective of extending this work to local-atomic dependent properties of other multicomponent materials, such as HEAs, the application is relatively straightforward. The primary modification required is to extend the definition of $R_{\bm{\phi}}$, $Q_{\bm{\theta}}$, and $P(\bm{y})$ distributions to include the additional chemical species involved.

%% tableau avec booktabs (sans filets verticaux)
\begin{table*}[p]
    \renewcommand{\arraystretch}{1.2}
    \renewcommand{\tabcolsep}{5pt}
    \vspace{0.5cm}
    \small
    \caption[Concentration of BSD defects $C_{\mathrm{BSD}}(T)$ per cationic site for different Pu concentrations, temperatures and BSD types, computed with IVAE models]{Concentration of BSD defects $C_{\mathrm{BSD}}(T)$ per cationic site for Pu concentrations: 10\%, 25\%, 50\%, 75\%, 90\%, temperatures: 500 K, 700 K, 900 K, 1100 K, 1300 K, 1500 K and 3 types of BSD. Each prediction is given by a separate IVAE model: 90 models were applied in total. All models, except those for T = 500 K, were tasked to generate 2000 4nn configurations during training, 50 configurations per training step, with $|\bm{y}|$ = 8. Models for T = 500 K were tasked to generate 6000 configurations instead (see Fig.~\ref{fig:IVAE_train_time_50Pu}). The computed values of defect concentrations are given as a mean $\pm$ 2 times standard deviation, calculated from the 5 highest predictions of the IVAE models.}
    \begin{center}
        \begin{tabular}{@{}l@{\hspace{35pt}}l@{\hspace{35pt}}l@{\hspace{35pt}}l@{\hspace{0pt}}} \toprule
        \toprule
        Environment & BSD1 & BSD2 & BSD3\\ \midrule
         $y_{\mathrm{Pu}}=10\%$, 500 K  & $(1.80 \pm 0.44)\times10^{-62}$ & $(1.87 \pm 0.74)\times10^{-53}$ & $(3.93 \pm 1.44)\times10^{-51}$ \\
         $y_{\mathrm{Pu}}=10\%$, 700 K  & $(6.20 \pm 1.39)\times10^{-45}$ & $(1.69 \pm 0.13)\times10^{-38}$ & $(8.71 \pm 0.68)\times10^{-37}$ \\
         $y_{\mathrm{Pu}}=10\%$, 900 K  & $(4.05 \pm 0.80)\times10^{-35}$ & $(3.91 \pm 0.34)\times10^{-30}$ & $(8.08 \pm 2.38)\times10^{-29}$ \\
         $y_{\mathrm{Pu}}=10\%$, 1100 K  & $(6.62 \pm 0.81)\times10^{-29}$ & $(8.10 \pm 0.74)\times10^{-25}$ & $(9.70 \pm 2.21)\times10^{-24}$ \\
         $y_{\mathrm{Pu}}=10\%$, 1300 K  & $(1.31 \pm 0.12)\times10^{-24}$ & $(4.03 \pm 0.72)\times10^{-21}$ & $(3.14 \pm 0.39)\times10^{-20}$ \\
         $y_{\mathrm{Pu}}=10\%$, 1500 K  & $(2.24 \pm 0.71)\times10^{-21}$ & $(2.25 \pm 0.22)\times10^{-18}$ & $(1.29 \pm 0.26)\times10^{-17}$ \\
         $y_{\mathrm{Pu}}=25\%$, 500 K  & $(5.87 \pm 2.36)\times10^{-62}$ & $(4.51 \pm 0.81)\times10^{-53}$ & $(1.17 \pm 0.26)\times10^{-50}$ \\
         $y_{\mathrm{Pu}}=25\%$, 700 K  & $(1.32 \pm 0.43)\times10^{-44}$ & $(3.34 \pm 0.50)\times10^{-38}$ & $(1.52 \pm 0.08)\times10^{-36}$ \\
         $y_{\mathrm{Pu}}=25\%$, 900 K  & $(8.19 \pm 3.34)\times10^{-35}$ & $(6.69 \pm 2.66)\times10^{-30}$ & $(1.34 \pm 0.65)\times10^{-28}$ \\
         $y_{\mathrm{Pu}}=25\%$, 1100 K  & $(9.16 \pm 3.03)\times10^{-29}$ & $(1.19 \pm 0.15)\times10^{-24}$ & $(1.42 \pm 0.21)\times10^{-23}$ \\
         $y_{\mathrm{Pu}}=25\%$, 1300 K  & $(1.80 \pm 0.46)\times10^{-24}$ & $(6.22 \pm 1.36)\times10^{-21}$ & $(4.39 \pm 2.00)\times10^{-20}$ \\
         $y_{\mathrm{Pu}}=25\%$, 1500 K  & $(2.60 \pm 0.16)\times10^{-21}$ & $(2.68 \pm 0.45)\times10^{-18}$ & $(1.75 \pm 0.35)\times10^{-17}$ \\
         $y_{\mathrm{Pu}}=50\%$, 500 K  & $(3.19 \pm 0.58)\times10^{-61}$ & $(2.49 \pm 0.27)\times10^{-52}$ & $(6.57 \pm 1.98)\times10^{-50}$ \\
         $y_{\mathrm{Pu}}=50\%$, 700 K  & $(3.77 \pm 0.50)\times10^{-44}$ & $(1.36 \pm 0.63)\times10^{-37}$ & $(4.74 \pm 0.83)\times10^{-36}$ \\
         $y_{\mathrm{Pu}}=50\%$, 900 K  & $(1.60 \pm 0.53)\times10^{-34}$ & $(1.55 \pm 0.60)\times10^{-29}$ & $(3.44 \pm 0.87)\times10^{-28}$ \\
         $y_{\mathrm{Pu}}=50\%$, 1100 K  & $(1.66 \pm 0.26)\times10^{-28}$ & $(2.41 \pm 0.54)\times10^{-24}$ & $(2.97 \pm 0.82)\times10^{-23}$ \\
         $y_{\mathrm{Pu}}=50\%$, 1300 K  & $(3.32 \pm 0.80)\times10^{-24}$ & $(9.57 \pm 1.96)\times10^{-21}$ & $(7.77 \pm 2.50)\times10^{-20}$ \\
         $y_{\mathrm{Pu}}=50\%$, 1500 K  & $(4.50 \pm 0.98)\times10^{-21}$ & $(4.44 \pm 0.77)\times10^{-18}$ & $(2.48 \pm 0.27)\times10^{-17}$ \\
         $y_{\mathrm{Pu}}=75\%$, 500 K  & $(7.77 \pm 3.32)\times10^{-61}$ & $(1.33 \pm 0.27)\times10^{-51}$ & $(2.58 \pm 0.97)\times10^{-49}$ \\
         $y_{\mathrm{Pu}}=75\%$, 700 K  & $(8.72 \pm 3.55)\times10^{-44}$ & $(2.73 \pm 0.84)\times10^{-37}$ & $(1.40 \pm 0.80)\times10^{-35}$ \\
         $y_{\mathrm{Pu}}=75\%$, 900 K  & $(3.08 \pm 1.24)\times10^{-34}$ & $(3.63 \pm 0.32)\times10^{-29}$ & $(6.61 \pm 0.14)\times10^{-28}$ \\
         $y_{\mathrm{Pu}}=75\%$, 1100 K  & $(3.77 \pm 0.64)\times10^{-28}$ & $(5.13 \pm 0.25)\times10^{-24}$ & $(5.33 \pm 2.06)\times10^{-23}$ \\
         $y_{\mathrm{Pu}}=75\%$, 1300 K  & $(5.44 \pm 1.23)\times10^{-24}$ & $(1.71 \pm 0.40)\times10^{-20}$ & $(1.25 \pm 0.23)\times10^{-19}$ \\
         $y_{\mathrm{Pu}}=75\%$, 1500 K  & $(7.60 \pm 2.53)\times10^{-21}$ & $(7.32 \pm 0.79)\times10^{-18}$ & $(4.39 \pm 1.13)\times10^{-17}$ \\
         $y_{\mathrm{Pu}}=90\%$, 500 K  & $(1.56 \pm 0.43)\times10^{-60}$ & $(2.36 \pm 0.54)\times10^{-51}$ & $(4.80 \pm 1.07)\times10^{-49}$ \\
         $y_{\mathrm{Pu}}=90\%$, 700 K  & $(1.68 \pm 0.25)\times10^{-43}$ & $(5.24 \pm 0.46)\times10^{-37}$ & $(2.15 \pm 0.15)\times10^{-35}$ \\
         $y_{\mathrm{Pu}}=90\%$, 900 K  & $(4.10 \pm 0.22)\times10^{-34}$ & $(5.94 \pm 1.18)\times10^{-29}$ & $(1.09 \pm 0.33)\times10^{-27}$ \\
         $y_{\mathrm{Pu}}=90\%$, 1100 K  & $(4.91 \pm 1.28)\times10^{-28}$ & $(7.70 \pm 0.83)\times10^{-24}$ & $(7.84 \pm 1.07)\times10^{-23}$ \\
         $y_{\mathrm{Pu}}=90\%$, 1300 K  & $(7.15 \pm 0.73)\times10^{-24}$ & $(2.56 \pm 0.70)\times10^{-20}$ & $(1.84 \pm 0.26)\times10^{-19}$ \\
         $y_{\mathrm{Pu}}=90\%$, 1500 K  & $(8.62 \pm 0.68)\times10^{-21}$ & $(9.97 \pm 2.65)\times10^{-18}$ & $(6.27 \pm 1.01)\times10^{-17}$ \\
        \bottomrule
        \bottomrule
        \end{tabular}
    \end{center}
    \label{tab:IVAE_cbsd_table}
\end{table*}

%% tableau avec booktabs (sans filets verticaux)
\begin{table*}[h!]
    \renewcommand{\arraystretch}{1.2}
    \renewcommand{\tabcolsep}{5pt}
    \vspace{0.5cm}
    \small
    \caption[Effective formation energy for different Pu concentrations, temperatures, and BSD types, computed from the predictions of IVAE models]{Effective formation energy, $E^\mathrm{f}_\mathrm{eff}$ [eV], for Pu concentrations: 10\%, 25\%, 50\%, 75\%, 90\%, temperatures: 500 K, 700 K, 900 K, 1100 K, 1300 K, 1500 K and 3 types of BSD. Each value was computed using the values of defect concentration predicted by IVAE models from Table~\ref{tab:IVAE_cbsd_table}, using Eq.~(\ref{eq:effEf})). All models, except those for T = 500 K, were tasked to generate 2000 4nn configurations during training, 50 configurations per training step, with $|\bm{y}|$ = 8. Models for T = 500 K were tasked to generate 6000 configurations instead (see Fig.~\ref{fig:IVAE_train_time_50Pu}). The computed values of effective formation energies are given as a mean $\pm$ 2 times standard deviation, calculated from the 5 highest predictions of the IVAE models.}

    \begin{center}
        \begin{tabular}{@{}l@{\hspace{77pt}}l@{\hspace{77pt}}l@{\hspace{77pt}}l@{\hspace{0pt}}} \toprule
        \toprule
        Environment & BSD1 & BSD2 & BSD3\\ \midrule
         $y_{\mathrm{Pu}}=10\%$, 500 K  & $6.13 \pm 0.01$ & $5.23 \pm 0.02$ & $5.00 \pm 0.01$ \\
         $y_{\mathrm{Pu}}=10\%$, 700 K  & $6.14 \pm 0.01$ & $5.25 \pm 0.01$ & $5.01 \pm 0.01$ \\
         $y_{\mathrm{Pu}}=10\%$, 900 K  & $6.14 \pm 0.01$ & $5.25 \pm 0.01$ & $5.02 \pm 0.02$ \\
         $y_{\mathrm{Pu}}=10\%$, 1100 K  & $6.15 \pm 0.01$ & $5.26 \pm 0.01$ & $5.02 \pm 0.02$ \\
         $y_{\mathrm{Pu}}=10\%$, 1300 K  & $6.16 \pm 0.01$ & $5.26 \pm 0.02$ & $5.03 \pm 0.01$ \\
         $y_{\mathrm{Pu}}=10\%$, 1500 K  & $6.15 \pm 0.04$ & $5.25 \pm 0.01$ & $5.03 \pm 0.02$ \\
         $y_{\mathrm{Pu}}=25\%$, 500 K  & $6.08 \pm 0.02$ & $5.19 \pm 0.01$ & $4.95 \pm 0.01$ \\
         $y_{\mathrm{Pu}}=25\%$, 700 K  & $6.10 \pm 0.02$ & $5.21 \pm 0.01$ & $4.97 \pm 0.01$ \\
         $y_{\mathrm{Pu}}=25\%$, 900 K  & $6.09 \pm 0.03$ & $5.21 \pm 0.03$ & $4.98 \pm 0.04$ \\
         $y_{\mathrm{Pu}}=25\%$, 1100 K  & $6.12 \pm 0.03$ & $5.22 \pm 0.01$ & $4.99 \pm 0.01$ \\
         $y_{\mathrm{Pu}}=25\%$, 1300 K  & $6.13 \pm 0.03$ & $5.21 \pm 0.02$ & $5.00 \pm 0.05$ \\
         $y_{\mathrm{Pu}}=25\%$, 1500 K  & $6.13 \pm 0.01$ & $5.23 \pm 0.02$ & $4.99 \pm 0.02$ \\
         $y_{\mathrm{Pu}}=50\%$, 500 K  & $6.00 \pm 0.01$ & $5.12 \pm 0.01$ & $4.88 \pm 0.01$ \\
         $y_{\mathrm{Pu}}=50\%$, 700 K  & $6.03 \pm 0.01$ & $5.12 \pm 0.03$ & $4.91 \pm 0.01$ \\
         $y_{\mathrm{Pu}}=50\%$, 900 K  & $6.04 \pm 0.03$ & $5.15 \pm 0.03$ & $4.90 \pm 0.02$ \\
         $y_{\mathrm{Pu}}=50\%$, 1100 K  & $6.06 \pm 0.01$ & $5.16 \pm 0.02$ & $4.92 \pm 0.02$ \\
         $y_{\mathrm{Pu}}=50\%$, 1300 K  & $6.06 \pm 0.03$ & $5.16 \pm 0.02$ & $4.93 \pm 0.04$ \\
         $y_{\mathrm{Pu}}=50\%$, 1500 K  & $6.06 \pm 0.03$ & $5.17 \pm 0.02$ & $4.94 \pm 0.01$ \\
         $y_{\mathrm{Pu}}=75\%$, 500 K  & $5.96 \pm 0.02$ & $5.05 \pm 0.01$ & $4.82 \pm 0.02$ \\
         $y_{\mathrm{Pu}}=75\%$, 700 K  & $5.98 \pm 0.02$ & $5.08 \pm 0.02$ & $4.84 \pm 0.03$ \\
         $y_{\mathrm{Pu}}=75\%$, 900 K  & $5.99 \pm 0.03$ & $5.08 \pm 0.01$ & $4.85 \pm 0.00$ \\
         $y_{\mathrm{Pu}}=75\%$, 1100 K  & $5.99 \pm 0.02$ & $5.08 \pm 0.01$ & $4.86 \pm 0.03$ \\
         $y_{\mathrm{Pu}}=75\%$, 1300 K  & $6.00 \pm 0.02$ & $5.10 \pm 0.02$ & $4.88 \pm 0.02$ \\
         $y_{\mathrm{Pu}}=75\%$, 1500 K  & $5.99 \pm 0.04$ & $5.10 \pm 0.01$ & $4.87 \pm 0.03$ \\
         $y_{\mathrm{Pu}}=90\%$, 500 K  & $5.93 \pm 0.01$ & $5.02 \pm 0.01$ & $4.79 \pm 0.01$ \\
         $y_{\mathrm{Pu}}=90\%$, 700 K  & $5.94 \pm 0.01$ & $5.04 \pm 0.01$ & $4.82 \pm 0.00$ \\
         $y_{\mathrm{Pu}}=90\%$, 900 K  & $5.96 \pm 0.01$ & $5.04 \pm 0.02$ & $4.82 \pm 0.02$ \\
         $y_{\mathrm{Pu}}=90\%$, 1100 K  & $5.96 \pm 0.03$ & $5.04 \pm 0.01$ & $4.83 \pm 0.01$ \\
         $y_{\mathrm{Pu}}=90\%$, 1300 K  & $5.97 \pm 0.01$ & $5.05 \pm 0.03$ & $4.83 \pm 0.02$ \\
         $y_{\mathrm{Pu}}=90\%$, 1500 K  & $5.97 \pm 0.01$ & $5.06 \pm 0.03$ & $4.82 \pm 0.02$ \\
        \bottomrule
        \bottomrule
        \end{tabular}
    \end{center}
    \label{tab:IVAE_eff_table}
\end{table*}

\clearpage

\section{Conclusions and perspectives}

We have demonstrated how Inverse Variational Autoencoder (IVAE) can be applied to investigate local-atomic dependent properties in chemically disordered compounds such as mixed-oxides $\mathrm{(U, Pu)O_2}$ nuclear fuels. IVAE allows us to both access those properties through the estimation of the partition function and sample the configurations from the configuration space

We have started with a mathematical implementation demonstrating how reverse Kullback-Leibler divergence can be applied in practice within a variational inference framework to estimate the partition function. In our study, based on Cooper-Rushton-Grimes (CRG) potential \cite{cooper2014many}, we showed that we can train IVAE in an unsupervised way to obtain accurate predictions of custom partition functions, which can be used for instance to compute concentrations of thermal defects. We have also shown a different application of IVAE, where we used its predictions to estimate the range of influence of local-atomic environments or average occupancies. All of those experiments were preceded by an accuracy and performance test, showing how the method performed in a controlled MOX environment. We have shown how the IVAE loss function can be modified to include the influence of global Pu concentration, and how to optimize its training performance. IVAE offers a versatile sampling framework applicable to various systems and objectives, allowing for detailed control and monitoring of the sampling process. The predicted probability distribution of configurations (but also any other part of the model) can be easily accessed and visualized, as we demonstrated, e.g., in Fig.~\ref{fig:IVAE_3D_10Pu}.

IVAE enabled us to overcome the limitations of the previously demonstrated Mixture Density Network (MDN) approach from \cite{karcz2023semi}. Namely, IVAE does not need any pre-prepared database for training, but it generates it on its own. Similarly, as with MDN, it also allowed us to access larger local environments that were beyond reach in the previous study \cite{didier2022disorder}. Unlike the MDN approach, IVAE also incorporates the effect of temperature (reflected by the $\frac{1}{k_{\mathrm{B}}T}$ factor, as we can see, for example, in Eq.~(\ref{eq:CdT_concentration_compressed})), which allows for richer physical insights on the computed properties. However, the advantage of the MDN method is that it requires fewer computational resources to train its models. MDN model required about 200 configurations, while in the case of IVAE, it is from 1000 to 6000 configurations. The increase in computational cost is a natural consequence of applying this type of network. In MDN we knew exactly which configurations to sample to train the model, while IVAE needed to find them on its own. It is important to note that this is a first-of-a-kind type of study involving a very novel type of machine learning architecture. There are a lot of places for improvement and further development of the method, that could reduce the computational time needed further. For example, IVAE performance could be improved by applying more advanced descriptors \cite{MiLADY, bartok2013representing, musil2021physics, goryaeva2021efficient}. Instead of generating atomic configurations, IVAE would predict the probability distribution of descriptors that could be then either transformed to atomic configurations directly (in the case of invertible descriptors) or be used to sample configurations from them.

In conclusion, the development and application of the IVAE models mark a significant step forward in the study of chemically disordered compounds. The flexibility and scalability of the IVAE framework, combined with its potential for further optimization, promise continued advancements in understanding and predicting properties in various chemically disordered systems. 

\section{Acknowledgments}

The authors thank T. Schuler and M. Nastar for the insightful discussions about the definition of defect properties in disordered compounds. This work was funded by the CEA FOCUS program "Expérimentation Numérique et Jumeau Numérique" (EJN) and contributes as well to the CEA SICOM/PICI2 project. It was provided with computing HPC and storage resources by GENCI at CINES and TGCC, thanks to the grant 2024-A0160906008 on the ROME partition of the Joliot Curie supercomputer and on the GENOA partition of Adastra.

\appendix

\section{\label{sec:AppendixIsing} Verification of IVAE model in Ising system}

We can verify the IVAE method by applying it to the Ising system and reproducing the results obtained in \cite{cantwell2022approximate}. Subsequently, we conduct an additional validation of the model's training performance. The Ising system is particularly suitable for our purposes for two reasons: it is extensively studied and well-documented in the literature, and it provides an easy transition to the system of atomic configurations in MOX. In the MOX environment, configurations of U and Pu atoms replace the up and down spins of the Ising system.

In our study, we will focus on the two-dimensional square lattice Ising system. We can define the energy function as:

\begin{equation}\label{eq:Ising_H}
    H(\bm{s}) = -J\sum_{\langle i, j \rangle} s^{(i)} s^{(j)} - h \sum_i^n s^{(i)},
\end{equation}

\noindent
where the spin at the site $i$ is denoted by $s^{(i)} \in \{ +1, -1 \}$, $J$ is the coupling strength, $\langle i, j \rangle$  denotes summation over nearest neighbors (each pair of sites is only included once in the sum), $n$ is the number of sites on the lattice and $h$ is an external magnetic field. In dimensions higher than 1, the Ising system has two distinct phases: a paramagnetic phase, where the spins are disordered due to thermal fluctuations, and a ferromagnetic phase where spins start to align in one direction. A phase transition separates those two phases at some critical temperature $T = T_c$, below which the system is ferromagnetic. To facilitate the comparison between the true partition function of the Ising system \cite{baxter2007exactly, moore2011nature} and the IVAE predictions, we assume $h = 0$ and $J = 1$. Under these conditions, the system's partition function can be computed analytically.

We can now define the partition function of the Ising system as:

\begin{equation}\label{eq:Ising_Z}
    Z = \sum_{i} f(\bm{s}^{(i)}),
\end{equation}

\noindent
with:

\begin{equation}\label{eq:Ising_f}
    f(\bm{s}) = e^{\beta \sum_{\langle i, j \rangle} s^{(i)} s^{(j)}},
\end{equation}

\noindent
where $\beta$ is the inverse temperature. The loss function of the IVAE model directly corresponds to estimating the true partition function of the system and is designed to be its lower bound. Therefore, we aim to maximize the value of the IVAE loss function. IVAE trains iteratively by generating a batch of spin configurations, computing their energy, and making estimations of the loss function as shown in Eq.~(\ref{eq:lnZ_fq_Jensen}), where we use $f(\bm{s})$ as a substitute of $f_T(\bm{x}_\mathrm{c})$. Initially, spins are chosen randomly. However, once the training is complete, the newly generated samples of configurations should approximately follow the true probability distribution of the system.

\begin{figure}[h!tbp]
    \vspace{0.5cm}
    \centering
    {\includegraphics[width=0.4\textwidth]{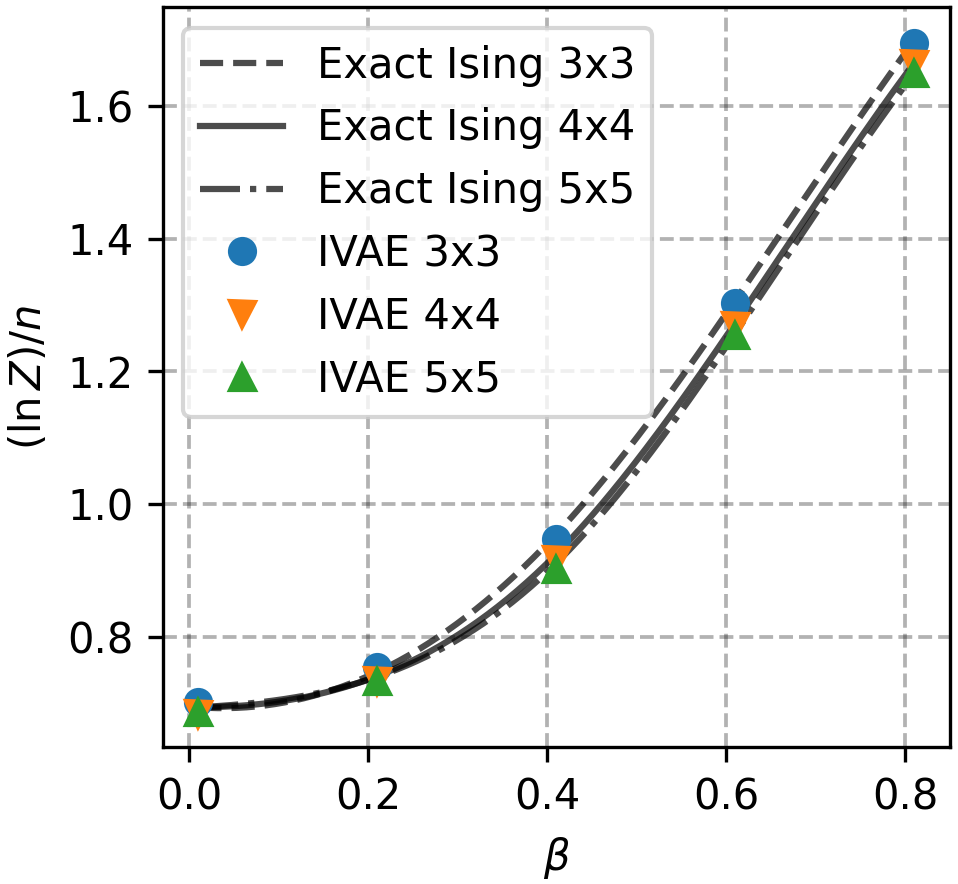}}
    \hspace{40pt}
    % First goest to the list of figures. Second is the description underneath
    \caption[Computing Ising system partition function with IVAE]{Computing Ising system partition function with IVAE. The lines show the exact computation, while the dots represent IVAE predictions in three different system sizes $n$ of 9, 16, and 25 spins.  The difference between model predictions and the exact computation is less than 1\%.}
    \label{fig:Ising_partition_function}
\end{figure}

The IVAE prediction results are shown in Fig.~\ref{fig:Ising_partition_function}. We considered three different Ising system sizes $n$ of 9, 16, and 25 spins, where we applied periodic boundary condition \cite{wu2014applying} to connect the left-most and right-most spins and similarly top with the bottom, to ensure the continuity of the spin distribution. The IVAE models used a simple neural networks architecture with a single hidden layer of 1024 neurons and the SELU\footnote{Scaled Exponential Linear Units (SELU) are one of the type of activation functions used in neural networks that induce self-normalizing properties in neural networks \cite{nguyen2021analysis}.} activation function, as described in \cite{cantwell2022approximate}. The temperature parameter of the Gumbel softmax function was set to $\tau = 1/10$.  All experiments used the Adam optimizer with TensorFlow 2.8 default parameters.

For the auxiliary distribution $P(\bm{y})$, we used a Bernoulli distribution. Due to the flexibility of the IVAE model, the size of the input vector  $\bm{y}$ does not need to match the size of the predicted Ising system, corresponding to the length of vector $\bm{\phi}$. However, the optimal size of $\bm{y}$ is not immediately clear. IVAE samples the Ising spin configurations by first sampling $\bm{y}$ from $P(\bm{y})$ and then mapping it to the spin configuration $\bm{s}$. The size of $\bm{y}$ affects both the expressiveness of the model and the computational cost. This is because larger sizes of $\bm{y}$ allow for more complex $\bm{y}$ to be transformed into $\bm{s}$, but they also necessitate more samples to sufficiently cover the corresponding configuration space of $\bm{y}$. Adequate sampling of the configuration space of $\bm{y}$ is crucial to accurately map the configuration space of $\bm{y}$ to the configuration space of $\bm{s}$. While the sampling cost for the Ising system is low (as each sample is a single summation over different spin pairs), it becomes significant for more complex systems like MOX, where each sample involves energy minimization with interatomic potentials or DFT, taking from minutes to thousands of CPU hours per sample. Ideally, $\bm{y}$ should be large enough to capture all key characteristics of $\bm{s}$, yet small enough to minimize computational cost. 

Since Eq.~(\ref{eq:lnZ_fq_Jensen_loss}) provides a lower bound, the goal is to find the $\bm{y}$ size that maximizes the loss function $-\Tilde{F}_{\bm{\phi}, \bm{\theta}}$. We tested different $\bm{y}$ sizes and found that $|\bm{y}|=8$ is sufficient for all Ising system sizes used in our experiments. Although the optimal $\bm{y}$ size might vary with system size and temperature, $|\bm{y}|=8$ resulted in an error of less than 1\% between IVAE predictions and the analytically computed Ising system partition function, without significantly increasing computational cost. Therefore $|\bm{y}|=8$  was used consistently across all Ising system experiments for all values of $\beta$.

\begin{figure}[h!tbp]
    \vspace{0.5cm}
    \centering
    {\includegraphics[width=0.4\textwidth]{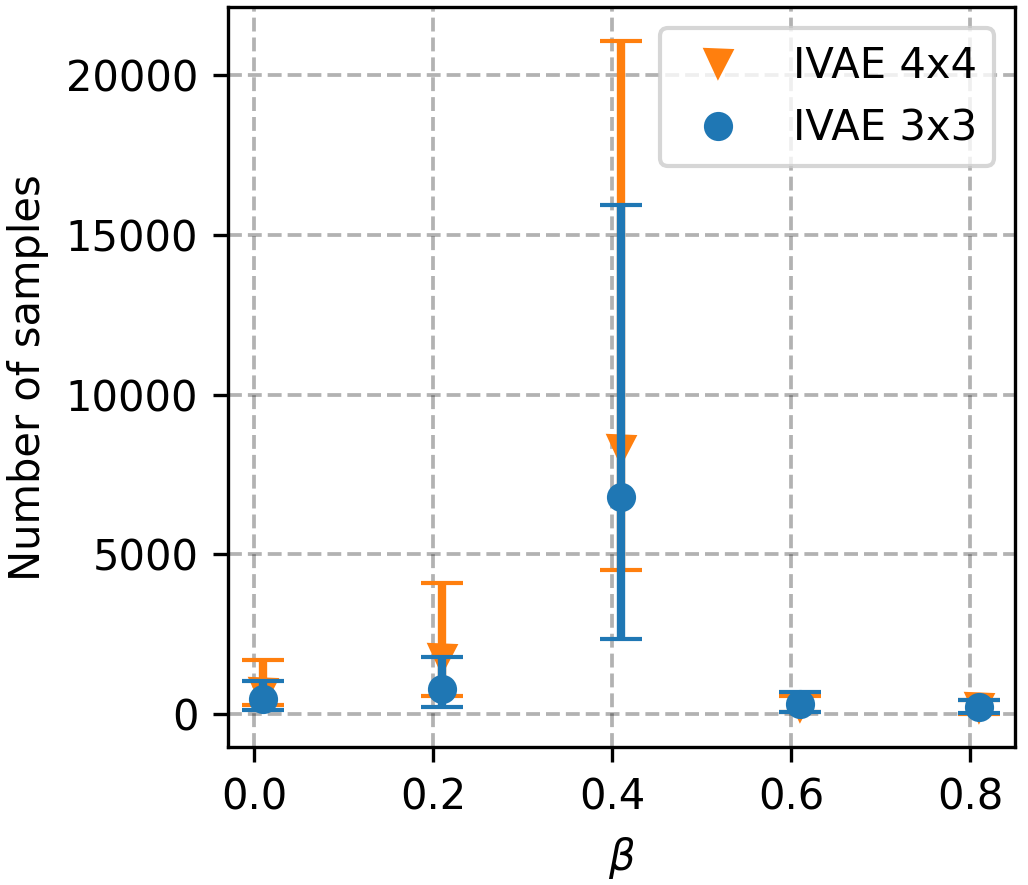}}
    \hspace{30pt}
    {\includegraphics[width=0.4\textwidth]{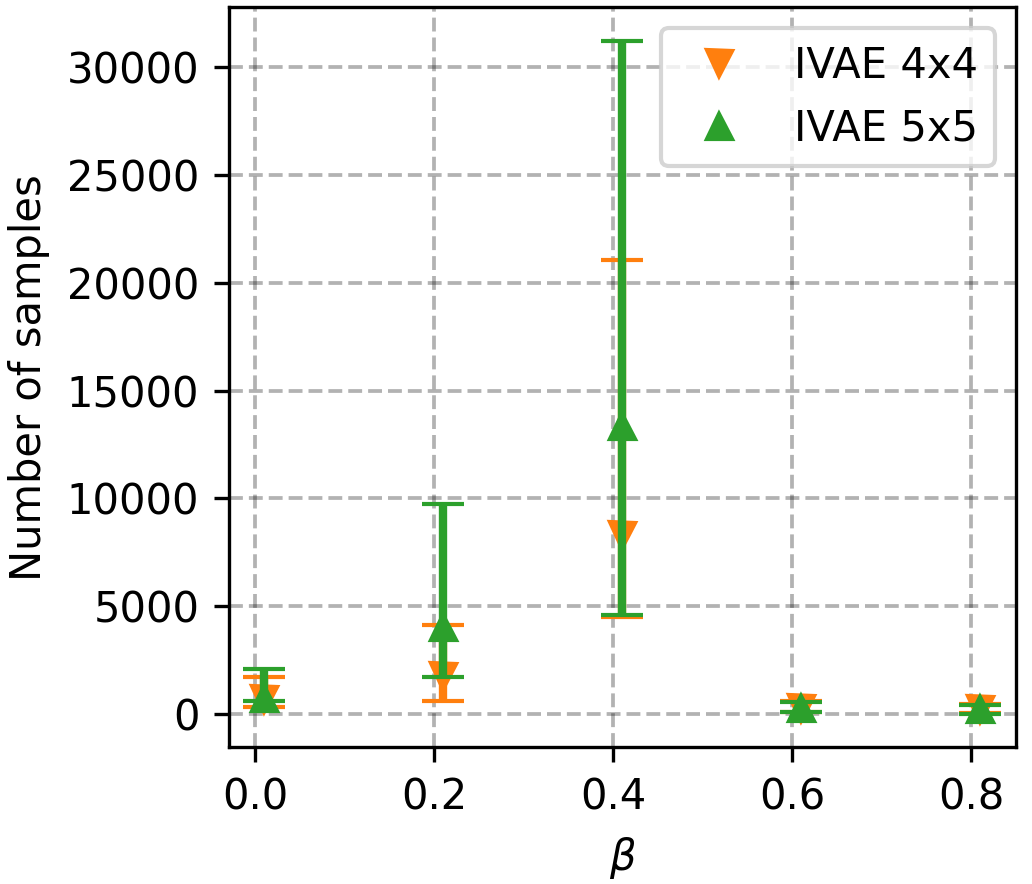}}
    % First goest to the list of figures. Second is the description underneath
    \caption[Accuracy of the IVAE predictions in the Ising system]{Accuracy of the IVAE predictions in the Ising system. We have measured the total amount of samples $\bm{s}$ needed to train the model to obtain an error of less than 1\% between IVAE predictions and the analytically computed Ising system partition function. The computed training cost is visualized as a 95\% confidence interval of the predictions of 60 IVAE models. The total number of possible configurations for 2D Ising systems with spins arranged in grids of sizes 3x3, 4x4, and 5x5 are 512, 65536, and 33554432, respectively.}
    \label{fig:Ising_accuracy}
\end{figure}

In Fig.~\ref{fig:Ising_accuracy} we can observe the training cost of the applied models, computed as the total amount of samples taken from the configuration space of Ising spins. There are two noticeable characteristics. The first one is that there is an increase in the training cost when we increase the size of the generated $\bm{s}$ configurations. This is not surprising, as a bigger system simply requires more exploration to cover its partition function. The second is that we can observe significant changes in the required training time depending on the temperature. This phenomenon occurs because the behavior of the Ising model changes significantly across different temperature ranges, particularly at the critical temperature $\beta_c = \frac{1}{2} \ln{1 + \sqrt{2}} = 0.4407$. At low temperatures, the system is in a stable, ordered state with most spins aligned. This results in a relatively simple configuration space, making it easier and faster for the machine learning model to learn and predict the system's behavior. Similarly, at high temperatures, the system is in a disordered state with spins flipping randomly, which again leads to a simpler learning task as the model can easily capture the random nature of the configurations.

However, at the critical temperature, the Ising model undergoes a phase transition from the ordered to the disordered state, resulting in more complex correlations between spins. This creates a highly intricate and varied configuration space that is, in a sense, in both phases simultaneously, making it much more challenging for the machine learning model to learn. As a result, the mean and variance of the measured training costs increase at the critical temperature. A possible remedy for this would be to use different forms for the $R_{\bm{\phi}}$ and $Q_{\bm{\theta}}$ distributions. However, such optimization could become quite challenging and might not necessarily be needed for our target application to the MOX system, so we did not explore it further.

\bibliographystyle{vancouver}
\bibliography{apssamp}% Produces the bibliography via BibTeX.

\end{document}